\documentclass[manuscript]{aastex}

\newif\ifAMStwofonts
\newcommand{\be}{\begin{equation}}
\newcommand{\ee}{\end{equation}}
\newcommand{\bea}{\begin{eqnarray}}
\newcommand{\eea}{\end{eqnarray}}

\begin{document}


\author{R.N. Henriksen}

\affil{Physics, Engineering Physics \& Astronomy,Queen's University} 
\affil{Kingston, Ontario, K7L 2T3, Canada}
\affil{henriksn@astro.queensu.ca}
\date{\today}



\begin{abstract}
  
 We begin by recalling the isothermal, collisionless, disc-halo. The disc component is the Mestel disc. Subsequently we introduce spiral arms to such an isothermal disc-halo system that are co-moving in the mean with an axi-symmetric background. These correspond to a similar disturbance in the halo, which is comprised of spiral structures on cones. The arms are necessarily transient due to the differential winding in the disc and their gradual destruction is described. Although the spiral potentials are weak compared to the axi-symmetric potential the arms are not propagating waves on the background, but rather co-move with it. They have an effect disproportionate to their relative magnitude on the gas distribution in the disc. The gas accumulates on the outside leading edge of the 'stellar' arm and an arm-interarm modulation of up to $100\%$ is possible. Compatible isothermal, scale-free, distribution functions are found either exactly or approximately for all of the collisionless  components of the disc-halo system. Repeated episodes of winding arms can produce an exponential disc.              
     
\end{abstract} 
\title{Transient Spiral Arms in  Isothermal Stellar Systems}

\setlength{\baselineskip}{13pt}
\section{Introduction}
\label{sec:intro}

Our objective in this paper is to construct an isothermal disc-halo and transient, isothermal, disc-halo spiral structure from a mixture of collisionless and gaseous matter. We restrict ourselves to an infinitely thin disc immersed in a background halo. The initial spiral arms are also infinitely thin in one approximation, where they are discrete. Both  artefacts may be regarded as the result of `coarse graining' the actual disc and arm. Thick discs merging smoothly into halo structures can be studied in the same fashion, but these have already resulted in the well known Evans models \cite{E93}. The formulation is meant to be gravitationally and dynamically self-consistent to a reasonable approximation (the co-moving circular particle velocity is small compared to the disc rotational velocity). 

 We begin by summarizing axi-symmetric, self-similar `isothermal disc-halos'\footnote{These are collisionless systems with similarity class $a= 1$.}. The discs do not have the same problems with gravitational equilibrium as do rigidly rotating discs and arms (e.g. \cite{Harxiv11}-similarity class $0$).   They may  require a compatible halo in order to remain stable to linear perturbations ( see \cite{ER98a}, \cite{ER98b} and \cite{GE99}).  In any case we do find a compatible halo in this paper  and together, the halo and the disc, define an axi-symmetric `isothermal disc-halo system'.
 
There is an infinitely large class (class $a\equiv\alpha/\delta$, a positive real number equal to the ratio of spatial ($1/\delta$) to temporal ($1/\alpha$) scales) of self-similar rotating thin discs, all of which possess differential rotation except class zero (rigid rotation). This differential motion presents the `winding problem' (\cite{BT08}) for non axially symmetric structures comprised of the same rotating material, which argument implies that such structures can not ultimately be stationary. 

This difficulty, which is common to all discs in differential rotation, has inspired a linear theory of spiral structure (\cite{LS66}) (see \cite{BT08} for a description of later developments). This theory derives the structure as a wave pattern propagating on the disc material. The pattern is assumed to be more nearly in rigid rotation with an angular velocity $\Omega_p$. However both simulations (\cite{JS2011}) and analysis (\cite{BT08}) suggest that these waves may also be transient.

This paper takes a rather different approach. The spiral arms are allowed to be material arms. They are normally transient and it is this evolution that we study in the non-linear limit. There is one case where the arms maybe in rigid rotation and long-lived, but after presenting the possibility we do not develop it further in view of the evidence.

 After summarizing the  axi-symmetric disc-halo structure that follows from isothermal collisionless matter, and discussing what rigidly rotating material arms would have to resemble, we model in detail collisionless spiral arms that are co-moving in the mean with the background disc rotational velocity. However these do not avoid the winding problem and are consequently transient. By focussing on the evolution of the spiral potential as the winding proceeds, we describe the gradual destruction by winding of the initially self-similar arms. The initial arms are maintained so long as the quantity $Vt/r$ ($V$ is the disc rotational velocity) is small. This restricts the lifetime at a given radius and the range of radii over which the arm persists at a given time. Time is to be measured from the establishment of the spiral structure which origin, either by instability or infall, we do not discuss.

The spiral disc potential must be associated with a non axially-symmetric potential component in the halo. The direction in which the causality operates may not always be the same, since the transient disc arms may be stimulated by the decaying orbit of an infalling object. In any case these spiral components are expected to be small compared to the axi-symmetric background potential. 

The various components of the disc-halo system are  constructed from a scale-free, isothermal distribution of collisionless particles plus scale-free, isothermal gas. Considerable discussion is given  to  the boundary condition on the potential at the disc. By using a distribution function approach we bypass solving for the detailed orbits of the arm particles. The orbits are nevertheless defined by the characteristics of the corresponding distribution function.    
 
The model presented is  not  a wave theory since the arms are comprised of a separate distribution of particles that rotates in the mean with the disc velocity. Recent simulations,(\cite{WBS11}), (\cite{KGC11}), and observations (\cite{FRDLW11}) encourage this point of view. The strength of the spiral potential  is  small compared to the total (disc plus halo) axi-symmetric potential, but it can nevertheless have a non-linear effect on the distribution of  gas in the disc. 

Although we do not solve for the gas dynamics consistently in this paper, it is likely that there is substantial streaming of gas and associated magnetic field through the arms. Such streaming can lead to shocks and hydraulic jumps (\cite{MC98}) in the gaseous matter. The magnetic field is essential to the full understanding of the gas dynamics.

 In the next section 2 we derive the isothermal disc-halo solution in axial symmetry. This represents the background for the non axially symmetric, isothermal structure. In section 3 we discuss the non axially symmetric disc and halo components, including their potential and distribution functions. Section 4 constructs an example of the disc-halo system with spiral structure. The final section is reserved for discussion and conclusions. 

\renewcommand{\textfraction}{0}
\renewcommand{\topfraction}{1} 
\renewcommand{\bottomfraction}{1}

\section{Axially Symmetric  Discs and Halos}\label{sec:a1axdynamics}
\subsection{Discs}

It is convenient to describe the familiar axi-symmetric Mestel disc in a differentially rotating reference frame. Without axial symmetry such a trick does not work because of the winding problem, but in this axi-symmetric example the self-similar analysis in such a frame allows us to establish a certain uniqueness for the self-consistent distribution function (DF). Hence the angular velocity of the locally rotating frame is $\Omega=V/r$.

We do not normally regard the Mestel disc as being an example of self-similarity, but in fact it is an isothermal example.
The self-similar surface density  is uniquely $\sigma=\Sigma/(\delta r)$ ($\Sigma$ is constant, $\delta$ may be thought of as an inverse arbitrary radius to appear more explicitly below), and this yields the corresponding potential above and on the disc, due to the disc, as (in cylindrical coordinates)
\be
\Phi_d=\frac{2\pi G\Sigma}{\delta}arcsinh(\frac{z}{r})+\frac{2\pi G\Sigma}{\delta}\ln{\delta r}.\label{eq:axiMpot}
\ee

We obtain this expression in spherical coordinates by letting $z\leftarrow r\cos{\theta}$ and $r\leftarrow r\sin{\theta}$. In either expression $\nabla^2\Phi_d=0$  above the disc and the boundary condition $1/(2\pi G)\partial_z\Phi_d=\sigma$ (or equivalently $-1/(2\pi G r)\partial_\theta\Phi_d=\sigma$) is satisfied at the disc. 

The final equation that determines self-consistency is 
\be
\sigma=\int~\int~F~dv_rdv_\phi,\label{eq:selfcon1}
\ee
where an appropriate two dimensional distribution function (DF) must be found. 

 A self-similar disc that depends on a constant velocity $V$ (or equivalently on a constant specific energy $E_V$) falls into the self-similarity class $a=1$ (\cite{CH91}). That is, the temporal scaling $1/\alpha$ is equal to the spatial scaling $1/\delta$ so that any velocity is not in fact scaled. Thus a constant velocity, or equivalently a constant specific energy, is compatible with this class of self-similarity. This constant may be used if necessary, to set finite  limits to the integration over the distribution function.

By enforcing rigorous self-similarity and a steady state in axial symmetry it can be shown (see e.g. \cite{HW95} for similar methods and also later sections f this paper) a general form of the DF is  

\be
F=K(C)e^{-(E_d'+Vv_\phi)/\Phi_o)},\label{eq:ssaxphysDF}
\ee
where $E_d'\equiv T+\Phi_o\delta R-V^2/2$ the `energy' in the locally co-moving frame. A constant $e^{(V^2/(2\Phi_o))}$ has been absorbed in $K(C)$. The one integral that preserves the self-similarity is found to be 
\be
C=(V+v_\phi)e^{-(T+Vv_\phi)},\label{eq:C} 
\ee
and $T=(v_r^2+v_\phi^2)/2$ in the locally rotating frame.

One readily finds that $E_d'+Vv_\phi=E_d$, where $E_d$ is the energy in the inertial frame. That is $E_d=E_d'+\omega rv_\phi$, which is familiar as the Jacobi integral.  

We note also from the form of $E_d$ that,if $v_\phi<V$, one can drop the term $Vv_\phi$ in equation (\ref{eq:ssaxphysDF}). This requires $\Phi_o>V^2$ if the bulk of the particles are to satisfy this condition. Subsequently $\Phi_o$ becomes $\Phi_{oa}$, which is due to the halo as well as the disc. Using this approximation with $K$ constant the DF has the isothermal form 
\be
F(E'_d)=\frac{\delta}{4\pi^2G}\exp{-(\frac{\delta E'_d}{2\pi G\Sigma})},\label{eq:isothermdiscDF}
\ee
 but with the energy in the locally rotating frame. The integral over velocities continues to give the necessary $1/r$ surface density, and the DF in terms of $E'_d$ is compatible with a mean rotation.  

The function $K(C)$ can be arbitrary (we have taken it to be a constant consistent with $\sigma=\Sigma/\delta r$ in the approximate argument above) , since equation (\ref{eq:selfcon1}) when the integral exists, will always  give  $\sigma\propto 1/r$. However in general the integral in this equation will yield for $\Sigma$ a complicated function of $\Phi_o$, $V$ and any amplitude constant appearing in the function $K(C)$. For example one might choose $K(C)=K_1\ln{(|C|+K_2)}$, whereupon with $K_1$, $K_2$ fixed and positive (with $|C|+K_2>1$) $\sigma$ may be calculated in principle given $V$ and $\Phi_o$. Such a DF is different from that often used in this context \cite{BT08}, so that even with strict self-similarity there is no absolute uniqueness. 

However this apparent generality is spurious if instead $\sigma$, $\Phi_o$ and $V$ are all fixed, since then an arbitrary choice of $K(C)$ ends by defining the associated amplitude constant in an ever more complicated way (other constants such as $K_2$ in the example above may be chosen for regularity of the DF).  It suffices then to make a choice for $K(C)$ that allows the chosen values, although this is clearly not a unique choice. A power law in the form $K(C)=K_dV^q/C^q$ serves this purpose and corresponds to one intuitively composed from the energy and angular momentum integrals (\cite{BT08}). The constant $K_d$ is a new fiducial constant with the dimensions of $F$, and it transpires subsequently that $q$ can be any  real number smaller than $1$.

We find thus a strictly self-similar  DF for the Mestel disc in the locally rotating frame according to (\ref{eq:ssaxphysDF}) as
\bea
F&=&\frac{K_d}{(1+\frac{v_\phi}{V})^q}\exp{\left(\frac{q(T+Vv_\phi)}{\Phi_o}\right)}\nonumber\\
&\times&\exp{\left(-(\frac{E'_d+Vv_\phi}{\Phi_o})\right)},\label{eq:Fpower}
\eea
where once again $e^{(V^2/2)}$ is absorbed into $K_d$.

We may  calculate $\sigma$ from equation (\ref{eq:selfcon1}) by integrating over velocities. Since the upper and lower limits in $v_r$ may be taken as positive and negative infinity respectively, we see that $q<1$ for the integral to converge. 

The lower limit in $v_\phi$ bears some thought.  A DF of the form ($\Theta(x)$ is the Heaviside function) $F=Pe^{-\delta R}$ where $P=\tilde P(v_r,v_\phi)\Theta(v_\phi+V)$ continues to satisfy the self-similar Boltzmann equation everywhere, but the same expression without $V$ does not. This means that the lower limit should be $v_\phi=-V$, when the angular momentum of this particle is zero. This implies that $C\ge 0$ for all particles in the ensemble.  

The integration over velocities for $\sigma$ now yields ($\Gamma(x)$ is the `gamma' or factorial function)
\be
\sigma=\frac{\sqrt{\pi}}{(1-q)\delta r}~\Gamma(\frac{1-q}{2})e^{A}K_d\Phi_o A^{q/2}, \label{eq:SSqsigma}
\ee
where 
\be
A\equiv \left(\frac{(1-q)V^2}{2\Phi_o}\right).\label{eq:A}
\ee

We recall that $\Phi_o\equiv (2\pi G\Sigma/\delta)$. If all particles were at rest in the rotating frame then  for equilibrium $\Phi_o=V^2$, if the disc is isolated. However this is not the case for collisionless particles as we calculate below. 

Equation (\ref{eq:SSqsigma}) is a relation between $K_d$, $\Phi_o$ and $\sigma$ plus $q$. To find the meaning of $q$ we calculate some mean quantities. One finds using the DF (\ref{eq:Fpower}) that $\overline{v_r^2}=\Phi_o/(1-q)$, or equivalently 
\be
q=1-\frac{\Phi_o}{\overline{v_r^2}},\label{eq:q}
\ee
and so the radial dispersion is greater than or less than $\Phi_o$ according as $q>0$ or $q<0$. 

The mean azimuthal velocity is easily found in the same way as 

\be
\overline{v_\phi}+V=\sqrt{\frac{2\Phi_o}{1-q}}~\frac{\Gamma(1-\frac{q}{2})}{\Gamma(\frac{1-q}{2})}.\label{eq:meanvphi}
\ee
If the mean velocity is taken to be  zero to enforce the net rotation, then setting the right-hand side of this last equation equal to $V$ yields the relation between $V$, $\Phi_o$ and $q$. One finds that $q<1$ for a reasonable result. To imitate the phenomenon of `asymmetric drift' (\cite{BT08}), one would have to allow $q$ and hence $\overline{v_r^2}$ to vary appropriately with radius.

The (squared) azimuthal velocity dispersion is simple in the inertial frame, taking the value $\overline{(v_\phi+V)^2}=\Phi_o$. From this result and equation (\ref{eq:meanvphi}) we find the squared dispersion in the locally rotating frame as
\be
\overline{v_\phi^2}=V^2\left(1+\frac{\Phi_o}{V^2}-2\frac{\Gamma(1-\frac{q}{2})}{A~\Gamma(\frac{1-q}{2})}\right).\label{eq:meanazdisp}
\ee

 These results parallel those in (\cite{BT08}), but we have shown that the DF may be found by requiring it to be scale-free rather than being an `ad hoc' function of energy and angular momentum. Moreover it follows from this treatment that although the DF of the self-similar Mestel disc is not unique, there is a restricted family of possible DF's. Each member of this family would give slightly different particle mean dynamics. We have chosen a sufficient DF that allows ready calculation and coincides with a previous choice. 

The effect of an isothermal spherical dark halo becomes clear after combining the argument above with that of the next section. Such a combination was studied long ago (e.g. \cite{MRS81},\cite{T82}), but we have derived it independently from our formal self-similar considerations.

\subsection{The Disc-Halo of Isothermal  Self-Similar Class}

We do not expect the halo above the disc to be in rotation, at least not with the amplitude of the disc rotation. We work on the halo therefore in this section in the inertial frame assuming any halo rotation to be small.
 
 A halo that is compatible with the Mestel disc will fall into the same self-similar class, but in three spatial dimensions. In three dimensions the velocity and the potential of this class are not scaled by powers of $r$ just as is the case for the disc, but the DF is scaled slightly differently as 
\be
f=P(\phi,\theta,\vec{v})e^{-2\delta R}.\label{eq:3DhaloDF}
\ee
The density scales similarly according to
\be
\rho=\Theta(\phi,\theta)e^{-2\delta R}.\label{eq:3Dhalodens}
\ee
Although the potential does not scale by a power of $r$, it is always possible when there is a constant velocity to include a logarithmic term so that the most general potential may be written as
\be
\Phi=\Phi_o\delta R+\Psi(\phi,\theta).\label{eq:edhalopot}
\ee
Here the constant $\Phi_o$ is compatible with the self-similarity  because it requires only a constant velocity squared for its dimension, while the logarithm ($\delta R$) is dimensionless.
 We use spherical polar coordinates and $R$ is once again the logarithmic radius according to $\delta r=e^{\delta R}$. 

With axial symmetry we can ignore the $\phi$ dependence in the potential, and so the Poisson equation becomes 
\be
\frac{4\pi G}{\delta^2}\Theta=\Phi_o+\frac{1}{\sin{\theta}}\frac{d}{d\theta}(\sin{\theta}\frac{d\Psi}{d\theta}).\label{eq:3Daxipot}
\ee

A spherically symmetric halo of this class gives an inverse square density law  (the singular isothermal sphere) according to $(4\pi G\Theta_s/\delta^2)=\Phi_{os}$ and $\rho=\Theta_s/(\delta r)^2$. With a Mestel disc added to this spherical halo the combined potential is $\Phi_c\equiv \Phi_d+\Phi_{os}\delta R$ (the disc potential is from equation (\ref{eq:axiMpot})), that is  
\bea
\Phi_c&=&\frac{2\pi G\Sigma}{\delta}\ln{\delta r}+\frac{4\pi G\Theta_s}{\delta^2}\ln{\delta r}\nonumber\\
&+&\frac{2\pi G \Sigma}{\delta}(\ln{\sin{\theta}}+arcsinh(\cot{\theta})).
\eea
We write this more simply as 
\bea
\Phi_c&=&\frac{2\pi G\Sigma}{\delta}\ln{\delta r}+\frac{4\pi G\Theta_s}{\delta^2}\ln{\delta r}\nonumber\\
&+&\frac{2\pi G \Sigma}{\delta}(\ln{(1+\cos{\theta})}).\label{eq:axiphitot}
\eea

This combined potential of the isothermal disc-halo satisfies the Poisson equation (\ref{eq:3Daxipot}) if $\Phi_o$ has the combined value 
\be
\Phi_{oc}\equiv (2\pi G/\delta)(\Sigma+2\Theta_s/\delta), \label{eq:combophi}
\ee
whence the potential at $\theta=\pi/2$ is $\Phi_{oc}\ln{\delta r}$. 
Normally in disc galaxies the second term in the potential is much larger than the first at an appropriate $\delta$, which justifies taking a spherically symmetric halo as a first approximation in this expression.

The question arises as to what type of matter forms the halo?  It might be isothermal gas or collisionless `isothermal' matter. However for galaxies it is of some interest to consider the constraints that follow from regarding it as comprised, at least in part, of collisionless matter. Since the halo density will now depend on the potential through the DF, a more complicated dependence on $\theta$ by the potential may be expected. 

We have analyzed  the general collisionless Boltzmann equation in spherical symmetry in the inertial frame. The rigorous \footnote{By `rigorous' we mean that we have disallowed any $R$ dependence in $P$ which leads to the unique self-similar isothermal DF.} application of $a=1$ self-similarity yields the unique `isothermal' distribution function as
\be
P=K_he^{-2{\cal E}_h/\Phi_o},\label{eq:isoP}
\ee
where ${\cal E}_h\equiv \Psi(\phi,\theta)+(\vec{v})^2/2$, and $\vec{v}$ is a three vector. When the scaling is applied to obtain the physical DF this becomes 
\be
f=K_he^{-2E_h/\Phi_o},\label{eq:isof}
\ee
where $E_h\equiv \Psi(\phi.\theta)+\Phi_o\delta R+(\vec{v})^2/2$. The DF retains this form even in the absence of any particular geometric symmetry. 

The density of such collisionless matter is given by 
 \be
\Theta_h=\int~P dv_r~dv_\theta~dv_\phi=(\pi \Phi_o)^{3/2}K_he^{-2\Psi/\Phi_o},\label{eq:isoSSdens}
\ee 
which must form at least part of the density appearing in the Poisson equation for the halo (\ref{eq:3Daxipot}) for self-consistency. In  axial symmetry this latter equation becomes ($\Phi_o\leftarrow \Phi_{oa}$)
\bea
\frac{4\pi G}{\delta^2}(\Theta_g&+&(\pi \Phi_{oa})^{3/2}K_{ha}e^{-2\Psi_a/\Phi_{oa}})\nonumber\\
&=&\Phi_{oa}+\frac{1}{\sin{\theta}}\frac{d}{d\theta}(\sin{\theta}\frac{d\Psi_a}{d\theta}),\label{eq:collaxipot}
\eea
where $\Theta_g$ represents the collisional, gaseous isothermal matter in the halo.
 
Isothermal gas in  static equilibrium satisfies $\Theta_g=\Theta_{gd}e^{-(2\Psi_a/\Phi_o)}$, where  the isothermal sound speed must be $c_s^2=\Phi_{oa}/2$ to be consistent with self-similarity. The constant $\Theta_{gd}$ is the value of $\Theta_g$ at the disc if we take $\Psi_a=0$ there.

It is  possible to solve equation (\ref{eq:collaxipot}) exactly for an axially symmetric disc-halo potential. We must use a disc boundary condition 
\be
-\frac{1}{2\pi G}\frac{d\Psi_a}{d\theta}=\frac{\Sigma_a}{\delta},\label{eq:boundary1}
\ee
 In addition we are free to set $\Psi_a(\pi/2)=0$. This allows the isothermal, scale-free, disc-halo system to be treated exactly. 

To obtain the solution for $\Psi_a(\theta)$ we introduce the `ad hoc' constant  
\be
Q\equiv \frac{4\pi^2 G}{\delta^2}(K_{ha}\sqrt{\pi\Phi_{oa}})(1+\frac{\Theta_{gd}}{(\pi\Phi_{oa})^{3/2}K_{ha}}),\label{eq:Q}
\ee
and then $y=\Psi_a/\Phi_{oa}$ in order to write equation (\ref{eq:collaxipot}) as 
\be
Qe^{-2y}=1+\frac{1}{\sin{\theta}}\frac{d}{d\theta}(\sin{\theta}\frac{dy}{d\theta}).
\ee

The solution follows by defining $y\equiv u+\ln{\sin{\theta}}$ since the resulting equation readily integrates for $u(\theta)$. We use the boundary conditions $u(\pi/2)=0$ and we impose the disc by $(du/d\theta)_{\pi/2}=-2\pi G\Sigma_a/(\delta\Phi_{oa})$. The solution that results for $\Psi_a$ is (a sign ambiguity is resolved by requiring $\Psi_a$ to be positive above the disc)
\bea
&~&e^{2\Psi_a/\Phi_{oa}}=\frac{Q}{Q+S}\sin^2{\theta}\times\nonumber\\
&~&\cosh^2{\left(C_2-\sqrt{Q+S}~\ln{\left(\frac{\sin{\theta}}{1+\cos{\theta}}\right)}\right)}.\label{eq:solisoDHg}
\eea
We have set 
\be
S=\frac{4\pi^2G^2\Sigma_a^2}{\delta^2\Phi_{oa}^2},\label{eq:S}
\ee
and 
\be
\cosh^2{(C_2)}=\frac{Q+S}{Q}.\label{eq:C2}
\ee

As $\theta\rightarrow 0$  equation (\ref{eq:solisoDH}) gives $(\Psi_a/\Phi_{oa}\approx((1-\sqrt{Q+S}))\ln{\theta}$. This goes to zero so that $(\partial_\theta\Psi)_0=0$ only if \footnote{I am obliged to the referee of a previous paper for calling my attention to this.} $Q+S=1$. Otherwise, provided that $Q+S>1$, the potential goes to positive infinity, the halo density goes to zero according to equation (\ref{eq:isoSSdens}), and $(\partial_\theta\Psi_a)_0\rightarrow -\infty$ as $-1/\theta$. This would require a negative mass per unit length on the axis and is unphysical by itself. 
The opposite case when $Q+S<1$ has the potential going to negative infinity on the axis and the density going to positive infinity there. The mass per unit length is then positive, which is also unphysical by itself. 

When $S+Q=1$ the expression (\ref{eq:solisoDHg}) simplifies substantially to 
\be
\frac{\Psi_a}{\Phi_{oa}}=\ln{(1+\sqrt{S}\cos{\theta})},\label{eq:solisoDH}
\ee
where we have added the subscript $'a'$ throughout this discussion to emphasize that this is an axi-symmetric disc-halo potential. This differs from the disc potential included in (\ref{eq:axiphitot}) only by the presence of $S$, which is indeed equal to unity for an isolated disc. The condition $S+Q=1$, is from the various definitions, a useful relation between the gas density plus collisionless density of the halo measured at the disc and the disc surface density. For negligible halo gas density this relation becomes
\be
\Sigma_a^2+\frac{\sqrt{\pi}K_{ha}}{4G}\Phi^{5/2}_{oa}=(\frac{\delta}{4\pi G})^2\Phi^2_{oa}.\label{eq:relation}
\ee 
According to this relation the assumption of isothermal self-similarity in the disc and halo implies a kind of disc-halo  'conspiracy', that is a smooth dominance transition in radius.
 
\begin{figure}
\begin{tabular}{cc} 
{\rotatebox{0}{\scalebox{.5} 
{\includegraphics{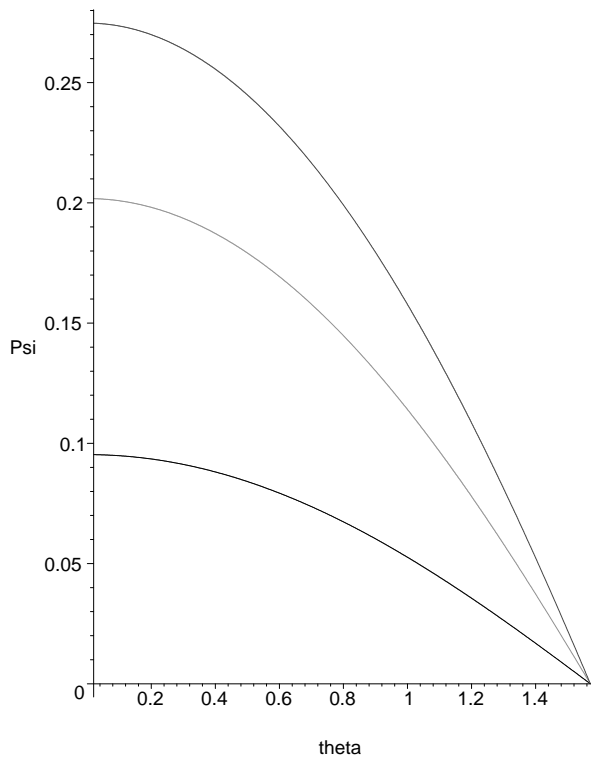}}}}
\end{tabular}
\caption{ The figure shows the variation of $\Psi$ in units of $\Phi_o$ in the range of $[0,\pi/2]$ radians. The parameters starting from the top curve are  $S=0.1,~.05,~.01$ respectively and $Q=1-S$ in each case. These conditions ensure a relatively unimportant disc in terms of mass and a relatively small component of isothermal gas in the halo.   }    
\label{fig:axiPsi}
\end{figure}

Examples of this  behaviour of the potential are shown in figure (\ref{fig:axiPsi}) when the disc and the isothermal gas are moderate perturbations to the isothermal collisionless halo. We see by considering the form for the density that as $S$ decreases, the density contours become more spherical.

This solution has been derived here in the context of our self-similar isothermal analysis, but it was known previously (\cite{MRS81}), (\cite{T82}), although without the isothermal halo gas contribution.

The distribution function of collisionless matter at the disc is now comprised of two components. From equations (\ref{eq:Fpower}) and (\ref{eq:isof}) we have in fact ($\delta_D(x)$ is the Dirac function)
\bea
&~&f=K_he^{(-2E_{ha}/\Phi_{oa})}+\frac{K_{da}\delta_D(v_\theta)}{(1+v_\phi/V)^q}\nonumber\\
&\times&\exp{\left((q-1)\frac{(E_{da}'+Vv_\phi)}{\Phi_{oa}}\right)}.\label{eq:comboDF}
\eea

In this expression $E_{ha}\equiv (v_r^2+(v_\phi+V)^2+v_\theta^2)/2+\Phi_{oa}\delta R+\Psi_a(\theta)$ for the halo population,  and $E_{da}'=(v_r^2+v_\phi^2)/2+\Phi_{oa}\delta R-V^2/2$ for the disc population recalling that $\Psi_a=0$. The velocities are relative to the locally rotating frame. The result is similar to the limiting DF found by Evans (\cite{E93}) when $q<0$.
 
Such a DF allows for various asymmetries in the stellar velocity ellipsoid at the disc.  
We note in particular that each of $E_{ha}$ and $E_d'+Vv_\phi$ are constants on their respective characteristic. If we follow a joint characteristic (traced by a fictitious particle) by holding the DF of equation (\ref{eq:comboDF}) constant then, provided that $v_\phi/V$ is small and/or $q\rightarrow 0$, we should expect each of these quantities to be constant on the joint characteristic defined by $f$ constant. Then taking the difference $E_{ha}-(E_d'+Vv_\phi)$ yields that the energy perpendicular to the disc 
\be
\frac{v_\theta^2}{2}+\Psi_a(\theta)\approx constant,\label{eq:Ez}
\ee    
along a fictitious characteristics close to that of the thin disc. This appears as a `third integral' in some disc models \cite{BT08}. In order for the fictious characteristic to be close to that of the disc, the disc population should dominate the halo population at the disc. The argument is even more direct if we neglect $v_\phi/V$ in the disc DF and take $q=0$ , so that the isothermal form $F=K_{da}e^{-E'_{da}}$applies also in the rotating disc.  

This isothermal disc-halo model is theoretically satisfying, but it suffers from two conflicts with observations. The first problem is that the strict self-similarity requires the disc to be infinite. However, a Mestel disc truncated at a radius $R_m$ has the potential
\be
\Phi=\frac{2\pi G\Sigma}{\delta}\ln{\frac{r}{4R_m}}
\ee
to first order in $r/R_m$ (the error is $O(r/R_m)^2$). Thus, to this order, the Mestel disc radial acceleration shares with that of the spherical halo  the property of depending only on the mass internal to radius $r$.  \footnote{This is not strictly true for the halo in the presence of the thin disc since then $\Psi_a=\Psi_a(\theta)$, but we may regard this as a small effect when $\Psi_a/\Phi_{oa}$ is small.} The isothermal disc-halo system can thus be regarded as forming the central part of some much more extended system.

The more serious clash with observations is that the mass surface density declines in radius as a power law and not as an exponential. The observed decline of the disc light is exponential on large scales (\cite{F70}) so that one must imagine much dark matter in the disc if the Mestel disc is to be taken seriously. This is not generally accepted for the following reasons. 

An isothermal HI disc that is supported by a mixture of rotation and pressure has a surface density that varies as $\Sigma_g(\delta r)^{(V^2-\Phi_{oa})/c_s^2}$. To obtain the self-similar $r$ dependence we must have therefore $\Phi_{oa}=V^2+c_s^2$. If such a compatible disc were sufficiently massive, it could provide the dark matter. However a typical HI surface density (\cite{BR97}) is $1M_\odot/pc^2$ which, if the solar neighbourhood is typical, is a negligible fraction of the disc mass (\cite{BT08}). The same conclusion applies to the total interstellar medium.

There is  moreover no dynamical evidence for dark matter in our galactic disc (\cite{E11}), so even a population of low mass stars and remnants (such as black holes and neutron stars) is excluded. In the end we are left again with the (steady/axi-symmetric) isothermal disc-halo system  being physically relevant only within one or two galactic scale lengths. Over this range it is possible that the discrepancy between the exponential and the power law is difficult to detect observationally. This is likely to be particularly true for Freeman type II spiral galaxies (\cite{F70}). The isothermal halo by itself appears to be  more widely applicable.
  
We do find one promising result in this regard in the succeeding sections, namely that wound up transient spiral structure acquires an oscillating exponential behaviour in the surface density. This takes the averaged form $\propto \exp{\sqrt{3}Vt/r}$ for a two-armed spiral at fixed $t$. Unfortunately this has a rather different shape from a pure exponential in radius, being `cuspier', and it only applies to the the spiral structure. Such structure is amplified through its effects on the gas however. 
  
We turn in the next section to study the transient spiral structures that may be imposed on an isothermal, axially symmetric, disc-halo background as reviewed above.
\section{Non Axially Symmetric Isothermal Disc-Halo}

\subsection{Steady, Rigidly Rotating Structure in the disc}
It may be that recurrent transient spiral structure in galaxies is the rule (e.g. \cite{JS2011}), and we shall study the non-linear temporal evolution of such arms in the next sub-section. However some recent studies (e.g. \cite{JS2012}) suggest that spiral structure may at least occasionally result from  growing instability to internal fluctuation. In such a self-excited, persistent mode,  the resulting arm should have a constant pattern angular speed $\Omega_p$ if it is to be long-lived when measured in galactic rotation periods. 

Such an arm will be a growing density wave in the background disc, but one expects the gravitational influence of  an eventual non-linear wave to modify the DF of the stellar disc. Thus even the nature of the Lindblad resonances that are so present in the linear theory may be modified (\cite{JS2012}). One way to describe the ultimate non-linear development of this process, is to assume the arm to be comprised of particles that have been entrained by the wave and move collectively with the constant pattern angular speed. 

The axi-symmetric Kalnajs disc (e.g. \cite{BT08}) is uniformly rotating, finite, and has unstable spiral modes especially when rapidly rotating. This suggests the constant generation of spiral density waves, but these are likely to be transient rather than steady. We would need  non-linear evolution into a rigidly rotating material spiral wave, in order to have long-lived structure.  

Because of its asymptotic nature and also because of the uniqueness it affords, we might assume that the DF of the entrained particles is compatible with rigidly rotating self-similarity. The similarity class of a spiral arm rotating with a non-zero, constant, $\Omega_p$ is $a=0$ rather than  the isothermal $a=1$ \cite{Harxiv11}. The scaling of the DF is the same in each class, namely $F=Pe^{-(\delta R)}$ but the scaling  of the surface density $\sigma=\Sigma e^{(\delta R)}$. The velocities are scaled in the $a=0$ class according to $\vec{v}=\vec{Y}e^{\delta R}$ with the consequent scaling of energy and potential. The logarithmic radius $R$ is unchanged from previous sections. By working in the rotating frame it was shown in \cite{Harxiv11} that the self-similar DF for a thin disc could be put in the  form 
\be 
F(E)=\frac{K}{\sqrt{|E'_d|}}.\label{eq:a0DF}
\ee
Once again the particle energy in the rotating frame is $E'_d\equiv (v_r^2+v_\phi^2)/2+\Phi_{eff}$, which is an integral of the particle motion. The effective potential is $\Phi_{eff}=\Phi-\Omega_p^2r^2/2$.

For strict self-similarity appropriate to rigid rotation, one must take $K$ constant and the upper limit in energy space $E_o$, either zero or $\propto \Omega^2r^2$. The scaled DF $P=\tilde P\Theta({\cal E}_o-{\cal E})$ ($\Theta$ is the Heaviside function) remains a solution of the CBE since the scaled energies are independent of $R$. The potential $\Phi$ is also proportional to $r^2$ in that case, and the surface density has the rather singular profile $\propto r$ rather than  $\propto r^{-1}$. 

The problem with such a material wave is that it must be strictly limited in radius and that the halo potential must adjust to stabilize the material with the self-similar form (\cite{Harxiv11}). The compatible halo is a core of uniform density as perturbed by the disc. In the presence of a background isothermal disc-halo the global self-similar 'conspiracy' would be broken. This does not seem like a successful model for large scale galactic spiral structure although it might describe nuclear structure. We continue to explore transient arms in this paper. 

\subsection{Transient, Corotating, Spiral Structure in the  Disc}

In this section we construct a non-axially symmetric, isothermal structure, that  rotates with a constant mean circular speed  $v_\phi=V$. We know that this can not be a steady configuration because  spiral structure  winds up  in time due to the differential rotation. For this reason we  treat the time dependence explicitly. We seek an approximate transient DF for the arms plus the details of how it is destroyed in time.

\begin{figure}
\begin{tabular}{cc} 
\rotatebox{0}{\scalebox{.4}
{\includegraphics{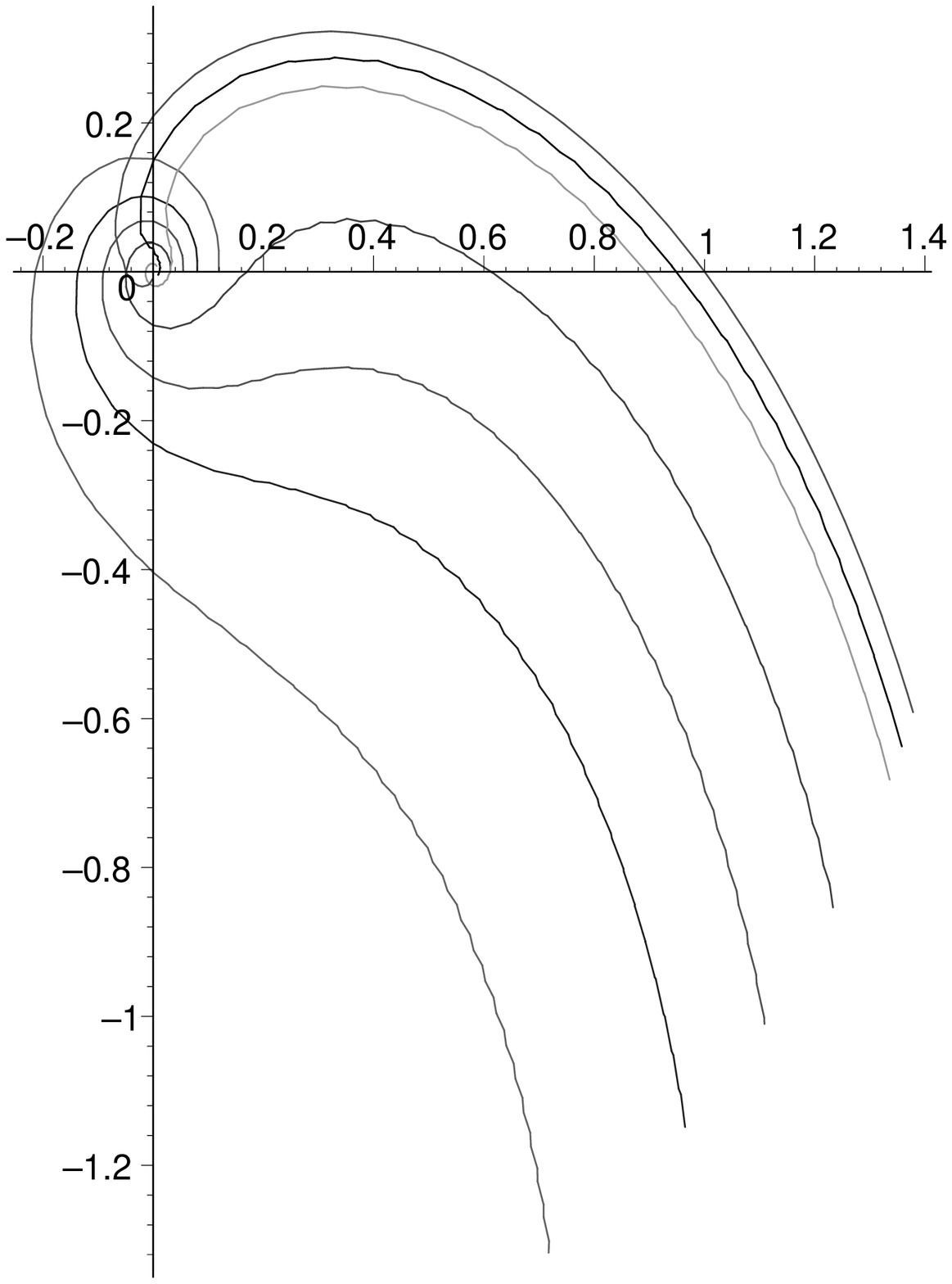}}}&
\rotatebox{0}{\scalebox{.4}
{\includegraphics{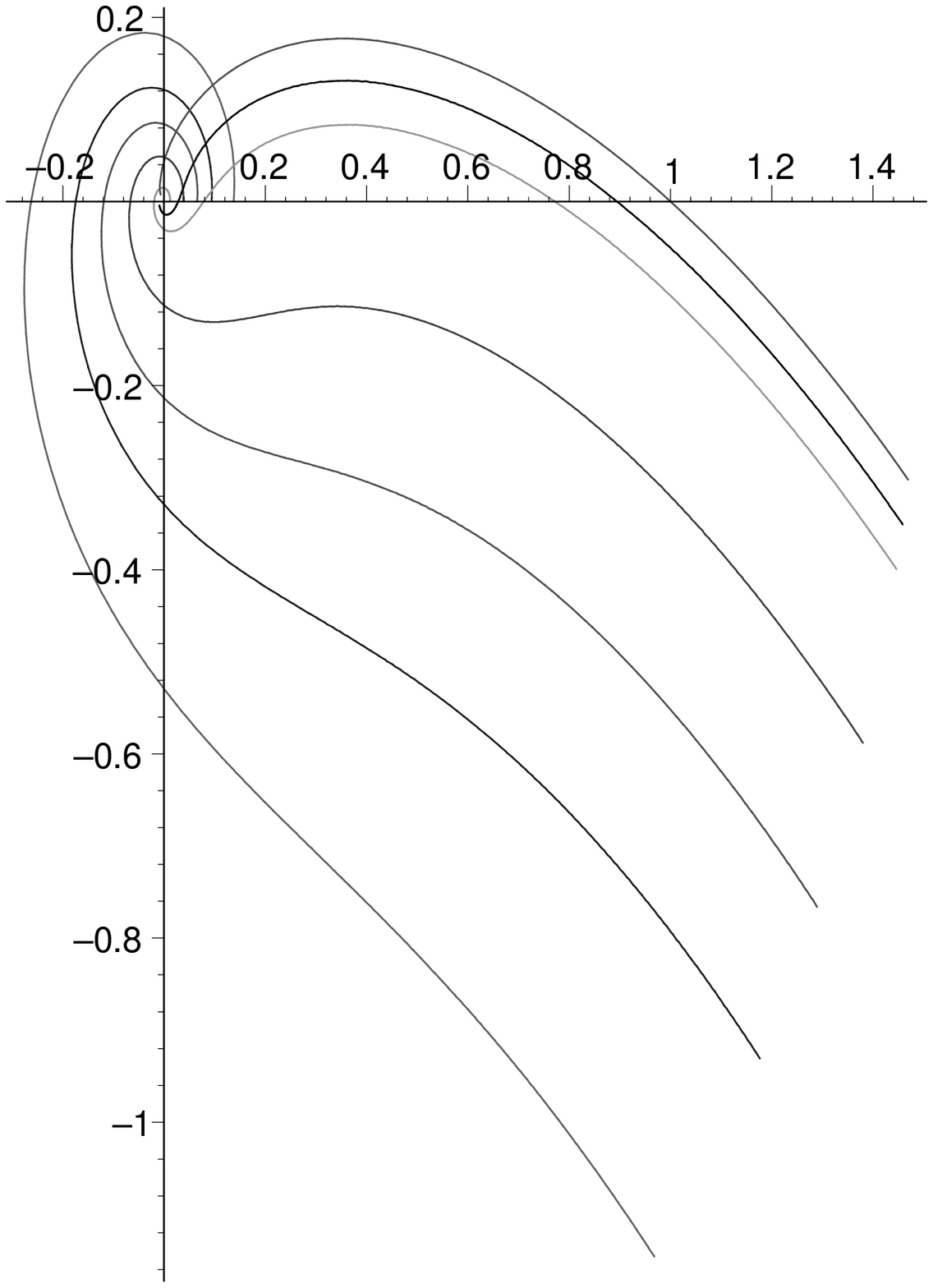}}}\\
\end{tabular}
\caption{The figure on the left shows the $\kappa=0$ spiral with initial winding angle $\epsilon/\delta=1$ at different times in the $xy$ plane. The figure on the right shows the same spiral  in the $xy$ plane with initial winding angle $\epsilon/\delta=0.5$ at the same times. The times are from the top curve to the bottom at $x=1$; $Vt=0,0.05,0.1,0.3,0.5,0.7,1.0$ respectively. The  initial spiral with smaller winding angle is distorted more rapidly. Distances are measured in terms of some fiducial radius $r_o$.  }    
\label{fig:windspiral}
\end{figure}

Such a model conceives the spiral structure to be `co-moving' with the axi-symmetric, inter-arm disc, rather than existing as a linear wave moving on the background. This description seems to correspond to the results of recent simulations reported in \cite{KGC11} and especially in \cite{WBS11}. The arms (both gaseous and stellar) found in these papers do mainly co-move during their transient existence. We do not suggest that all arms behave in this fashion, since we know that sufficiently small disturbances will propagate as waves on a background. In fact this model might be considered as a non-linear wave, since it  does `propagate' eventually due to winding (see figure \ref{fig:windspiral}). Moreover there is likely to be  relative motion between the interarm gas and these arms while they persist.  

We treat this problem by remaining close to a self-similar evolution in time, at least before major winding has occurred. The explicit CBE equation is the disc version of the equation studied later for the rotating component of the halo in the next sub-section. We have chosen a local frame  that is time independent and coincides with the velocity $V$ of the flat rotation curve of the background disc. Thus once again $\Omega=V/r$ and the relevant equation becomes 

\bea
&\partial_tF&+v_r\partial_r F+(\frac{v_\phi}{r}-tv_r\partial_r\Omega)\partial_\phi F\nonumber\\
&+&\!\!\!(\frac{v_\phi^2}{r}+2\Omega v_\phi+\Omega^2 r-\partial_r\Phi)\partial_{v_r}F\nonumber\\
&-&\!\!\!(\frac{v_\phi v_r}{r}+2\Omega v_r+v_r r\partial_r\Omega)\partial_{v_\phi}F=0.\label{eq:stac2be}
\eea  

The formal  procedure has been discussed  elsewhere (\cite{LeDHM11b}, and references therein) so we will only outline it here.
We use a logarithmic time $T$ as the self-similar Lie parameter and introduce on dimensional grounds  the scaled quantities $R$, $\vec{Y}$, $\xi$, $\Psi$ and $P$ according to  
\bea
&\delta t=e^{\delta T},r=Re^{\delta T},\xi=\phi+\epsilon T,\sigma=\Sigma e^{-\delta T}\nonumber\\ 
&F=P(R,\xi,v_r,v_\phi;T)e^{-\delta T},\vec{v}=\vec{Y},\nonumber\\
&\Phi_{dr}=\Phi^{(r)}_{do}\ln{(\delta R/V)}+\Phi^{(r)}_{do}\delta T+\nonumber\\
&\Psi_{dr}(R,\xi,;T).\label{eq:timedepvars}
\eea
Formally $\delta$ has the dimension of reciprocal time, but in fact all temporal and spatial quantities (and consequently velocities) may be thought of as numerical values in terms of some fiducial radius $r_o$ and fiducial time $t_o$.

The form of the potential  is equivalent to  
\be
\Phi_{dr}\equiv \Phi^{(r)}_{do}\ln{(\delta r/V)}+\Psi_{dr}(R,\xi,\theta),\label{eq:temppot}
\ee
and we recall that there is self-similarity in time only if $P$, $\Sigma$ and $\Psi_{dr}$ are independent of $T$. The winding term destroys this in a secular manner that we discuss below.
  
After writing the CBE in terms of these variables we obtain from it  in the usual way the characteristic equations
\bea
\frac{dP}{dT}&=&\delta P,\frac{dR}{dT}=Y_R-\delta R,\nonumber\\
\frac{d\xi}{dT}&=&\epsilon+\frac{Y_\phi}{R}+\left(\frac{V}{\delta R}\right)\frac{Y_R}{R},\label{eq:timedepchars}\\
\frac{dY_R}{dT}&=&\frac{Y_\phi^2}{R}+\frac{2VY_\phi}{R}+\frac{V^2}{R}-\frac{\Phi_{o}}{R}-\partial_R\Psi_{dr},\nonumber\\
\frac{dY_\phi}{dT}&=&-\frac{1}{R}\left(Y_\phi Y_R+VY_R+\partial_\xi\Psi_{dr}\right).\nonumber
\eea
We use $\vec{Y}$ to distinguish the scaled equations, but it is identical to $\vec{v}$.
 
The $Y_\phi$ characteristic equation may be combined with the characteristic expression for $dR/dT$ to give
\be
\frac{d}{dT}\left(\ln{((Y_\phi+V)Re^{\delta T})}\right)=-\frac{1}{R(V+Y_\phi)}\partial_\xi\Psi_{dr},\label{eq:scaleangmom}
\ee
which in physical variables is the angular momentum equation
\be
\frac{d}{dt}(r(v_\phi+V))=-\partial_\xi\Psi\equiv -\partial_\phi\Phi_{dr}.
\ee

The $R$, $\xi$ characteristics may be combined with the $Y_R$, $Y_\phi$ characteristics to obtain an energy equation in the form
\bea
&~&\frac{dE'_{dr}}{dT}=(\partial_T\Psi_{dr}-\delta R\partial_R\Psi_{dr}+\epsilon\partial_\xi\Psi_{dr})\nonumber\\
&+&V(V+v_\phi)\left(\frac{d}{dT}(\ln{Re^{\delta T}})\right)\nonumber\\
&+&\left(\frac{V}{\delta R}\right)\frac{Y_R}{R}\partial_\xi\Psi_{dr}.\label{eq:entemprot}
\eea
Here $E'_{dr}\equiv \vec{Y}^2/2+\Phi_{dr}$, where $\Phi_{dr}=\Phi^{(r)}_{do}\ln{\delta R/V}+\delta \Phi^{(r)}_{do}T+\Psi_{dr}$, is the energy in the locally rotating frame at the disc. 

We may eliminate $Y_R/R\equiv d(\ln{Re^{\delta T}})/dT$ between this energy equation and equation (\ref{eq:scaleangmom}) to obtain 
\bea
&~&\frac{dE'_{dr}}{dT}=(\partial_T\Psi_{dr}-\delta R\partial_R\Psi_{dr}+\epsilon\partial_\xi\Psi_{dr} )\nonumber\\
&+&\!\!\!\left(\frac{V}{\delta R}\right)(\frac{Y_R}{R}-\delta)\partial_\xi\Psi_{dr}-V\frac{d(V+Y_\phi)}{dT}.\label{eq:engtempinert}
\eea
We note that $E_{dr}=E'_{dr}+V(V+Y_\phi)$, which is the energy equal to $\Phi_{dr}+(Y_\phi+V)^2/2+Y_R^2/2$ in the inertial frame but for a constant $-V^2/2$. Thus the last equation can be written as $dE_{dr}/dT$ equal to the terms on the right that involve $\Psi_{dr}$. So long as the spiral structure remains self-similar, it will be steady in the locally rotating frame. We would like $E'_{dr}$ to be an integral of the particle motion during this phase, and this will be approximately the case if $Y_\phi<V$ and if $E_{dr}$ is an integral.

To obtain $E_{dr}$ as an integral we must set the right-hand side of equation (\ref{eq:engtempinert}) involving $\Psi_{dr}$ to zero. When used with the radial characteristic to eliminate $Y_R$, this is a linear equation for $\Psi_{dr}$. The general solution has the form
\bea
\Psi_{dr}&=&\Psi_{dr}(\xi-\epsilon T+V/(\delta R), Re^{\delta T})\nonumber\\
&\equiv&\Psi_{dr}(\phi+\Omega(r)t,r),\label{eq:Psidiscgen}
\eea
where $\phi+\Omega(r)t\equiv \phi_I$ and $\phi_I$ is the inertial frame angle.
This merely confirms that a steady potential is required to obtain a steady distribution in the inertial frame. An explicit dependence on $T$, or equivalently $r$ in the above expression breaks the self-similarity.

However we do not wish to describe material in the inertial frame. We can create a potential based on transient logarithmic spiral by taking  one variable to be a combination $\kappa$ of the above coordinates in the form 
\bea
\kappa&\equiv&\xi-\epsilon T+\frac{V}{\delta R}+(\frac{\epsilon}{\delta})\ln{(Re^{\delta T})}\nonumber\\
&\equiv& \xi+\frac{\epsilon}{\delta}\ln{R}+\frac{V}{\delta R}\nonumber\\
&\equiv& \phi+(\frac{\epsilon}{\delta})\ln{r}+\Omega(r)t.\label{eq:tempspiral}
\eea 
Here we have supposed that radii are in terms of a fiducial quantity that might be $r_o=V/\delta$, and we recall that $\Omega=V/r$.

We retain the other variable as $r=Re^{\delta T}$. Hence 
\be
\Psi_{dr}=\Psi_{dr}(\kappa,r),\label{eq:Psidiscspiral}
\ee
and this form must ultimately be made compatible with the Poisson equation. We observe once again that although the winding term $V/(\delta R)=\Omega(r)t$ is compatible with the self-similarity (requires no $T$ dependence), the dependence on $r=Re^{\delta T}$ is not. We shall see below that this dependence on $r$ is generally required in order to satisfy the Poisson equation in the presence of the winding term. Thus the similarity is broken by this effect, as might be expected.

We wish now to write a  DF for material at rest on average in the rotating frame, which is compatible with the rotating potential. We do not strictly have an integral in this frame since equation (\ref{eq:engtempinert}) is currently exact in the form 
\be
\frac{dE'_{dr}}{dT}=-V\frac{d(V+Y_\phi)}{dT},\label{eq:dErotdT}
\ee
which gives the integral $E_{dr}=E'_{dr}+V(V+Y_\phi)$ as discussed above. However it is clear that provided $Y_\phi/V$ in the local rotating frame is small, we may treat $E'_{dr}$ as an integral for those particles. This condition generally holds for the majority of particles in spiral galaxies.
 
 Since this is the only identified integral, we  write the characteristic solution of the Boltzmann equation as the approximate DF $P=F(E'_{dr})e^{\delta T}$. But $\Sigma=\int~P~dY_RdY_\phi$ and this should be independent of $T$ for self-similarity, to which behaviour we wish to remain as close as possible for uniqueness. Thus, recalling the form of the potential (e.g. see after equation (\ref{eq:entemprot})),  we see that we must have the isothermal DF in the locally rotating frame 
\be
F(E'_{dr})=K_{dr}\exp{\left(-\frac{E'_{dr}}{\Phi^{(r)}_{do}}\right)},\label{eq:DFtransspiral}
\ee
where $K_{dr}$ is the normalization for the transient spiral distribution function. In order for the bulk of the particles to obey the condition $Y_\phi<V$ we should require $\Phi^{(r)}_{do}< V^2$.  The mean velocity of these particles is zero in the locally rotating frame, due to the symmetry of the DF.
When other components are present that may be described by the isothermal DF, the potential in the exponential will be the sum of the various potentials.

This DF also gives  $\Sigma\propto 1/R$ and hence $\sigma=\Sigma e^{-\delta T}\propto 1/r$ as it should for self-similarity. However because of the likely dependence on $r$ in $\Psi_{dr}$ that we now pursue, this self-similar behaviour will be broken in general as the winding continues. This also leads to a more interesting radial variation of the spiral surface density, although it is transient. 

To obtain the equation for the disc potential above the plane we use the Poisson equation 
\bea
\!\!\!\!\!\!\!\!\!\!\!\!\!\!\!\!\!&~&\frac{1}{r^2}(\partial_r(r^2\partial_r(\Phi_{dr}))\nonumber\\
\!\!\!\!\!\!\!\!\!\!\!\!\!\!\!&+&\frac{1}{\sin^2{\theta}}\partial_\theta(\sin{\theta}\partial_\theta\Phi_{dr})+\frac{1}{\sin^2{\theta}}\partial^2_\phi~\Phi_{dr})=0,
\eea
and insert the spiral form    
\be
\Phi_{dr}=\Phi^{(r)}_{do}\ln{r}+\Psi_{dr}(\kappa,\theta,r),
\ee
to find eventually
\bea
&~&\Phi^{(r)}_{do}+\frac{\epsilon}{\delta}\partial_\kappa\Psi_{dr}+\partial_r(r^2\partial_r\Psi_{dr})\nonumber\\
&+&(\frac{\epsilon}{\delta}-\frac{Vt}{r})r\partial_r\partial_\kappa\Psi_{dr}\nonumber\\
&+&\left((\frac{\epsilon}{\delta}-\frac{Vt}{r})^2+\frac{1}{\sin^2{\theta}}\right)\partial^2_\kappa\Psi_{dr}\nonumber\\
&+&\frac{1}{\sin{\theta}}\partial_\theta(\sin{\theta}\partial_\theta\Psi_{dr})=0.\label{eq:rotPoiss}
\eea
For brevity subsequently we write the differential operator in this equation according to 
\be
{\cal L}\Psi_{dr}+\Phi^{(r)}_{do}=0,\label{eq:shortrotPoiss}
\ee
and $\Phi^{(r)}_{do}$ may be taken zero by absorbing the log potential into the corresponding axi-symmetric term, when present.

One can only neglect the $r$ dependence in this equation, and so preserve strict self-similarity, if $Vt/r\equiv \Omega(r)t<\epsilon/\delta$.  This might have been expected, but the dependence on the initial winding angle $\epsilon/\delta$ is of interest. In the circular arm initial limit ($\epsilon/\delta\rightarrow\infty$) the winding can, not surprisingly, be for an indefinite time.   

This linear equation is readily solved in terms of modes of the form
\be
\Psi_{dr}=\Phi^{(r)}_{do}\ln{\sin{\theta}}+e^{(im\kappa)}T(\theta){\cal R}(r),\label{eq:Psid}
\ee
where as already remarked $\Phi^{(r)}_{do}$ may be absorbed into $\Phi^{(a)}_{do}$ and so taken zero here. 
We recall that $\kappa=\phi+(\epsilon/\delta)\ln{r}+Vt/r$ and $\phi$ is in the locally rotating frame. The log spiral is completely wound up at a fixed $r$ when $\Omega(r) t=2\pi$. This gives $\approx 10^{7.5}$ years at $r=10$ kpc and $V=200$ km/sec. However there is an outward moving `winding wave', given by $Vt/r=cst<\epsilon/\delta<2\pi$, outside of which the log spiral remains recognizable and similarity is maintained.

The rotating DF and the rotating potential are linked through the disc boundary condition 
\be
2\pi G\sigma_{dr}=-\frac{1}{r}(\partial_\theta\Psi_{dr})|_{\theta=\pi/2},\label{eq:discsigma}
\ee   
where 
\be
\sigma_{dr}=\int~F(E'_d)~dY_RdY_\phi\equiv 2\pi \Phi^{(r)}_{do} K_{dr}\frac{e^{-\frac{\Psi_{dr}}{\Phi^{(r)}_{do}}}}{r}.
\ee
This condition is generally difficult to satisfy for all $\kappa$ for a single mode, because of the exponential dependence of $\sigma$ on the potential that follows from the last integral. Fortunately we can choose to satisfy it everywhere by adding isothermal gas to the disc that is not described by the isothermal collisionless DF. 
If however the arms are deemed to be predominantly comprised of collisionles matter, then we can  satisfy the boundary condition only at discrete values of $\kappa$. These then become the idealized spiral arms much in the same fashion that the razor-thin disc is an idealized thick disc. 

For a fixed spiral trajectory, the particles ought to be constrained to move one dimensionally along the spiral in the co-moving frame. However at fixed $\kappa=\phi+(\epsilon/\delta)\ln(r)+Vt/r\equiv \nu+Vt/r$, the log spiral sweeps over different trajectories according to $d\nu=-Vd(t/r)$. The effective arm therefore will be extended and distorted so we  continue to allow a two-dimensional DF to describe the arm particles. This is illustrated in  figure (\ref{fig:windspiral}). We see that the winding occurs early at small radii and later at larger radii. Moreover the arm appears to move as a  non-linear wave in the co-moving frame at large radii before it is  completely distorted, which one expects to thicken the actual arm. The slowest destruction occurs for the larger winding angle so that arms with large winding angles (including rings) are most likely to be observed.

This concludes our model for transient arms in an isothermal, thin, disc. In the next sub-sections we treat the compatible non axially symmetric halo and the necessary isothermal gas component.

\subsection{The  Transient Non-Axially Symmetric  Halo}

The total potential in a disc-halo system must satisfy the Poisson equation in the form
\be
\nabla^2\Phi=4\pi G\rho_h+4\pi G\Sigma\frac{\delta_D(\theta-\pi/2)}{r},\label{eq:PDH}
\ee
where the total potential is the sum of that due to the disc and that due to the halo namely $\Phi=\Phi_d+\Phi_h$. This equation separates for the two components to give
\bea
\nabla^2\Phi_h&=&4\pi G \rho_h,\nonumber\\
\nabla^2\Phi_d&=&0,\label{eq:PDHS}
\eea
where the total disc potential $\Phi_d$ satisfies the boundary condition (\ref{eq:discsigma}) when $\sigma$ is the total surface density.

Recalling the previous sections, we can form the potential due to the disc from two components. These are respectively the axi-symmetric potential $\Phi_{da}(\theta)$ given as the disc component in (\ref{eq:axiphitot}) plus the rotating structure $\Phi_{dr}=\Phi^{(r)}_{do}\ln{r}+\Psi_{dr}$. The potential $\Psi_{dr}$ is a solution of equation (\ref{eq:rotPoiss}). Thus we write ($\delta$ of the steady state spatial scaling is replaced by the time dependent scaling $\delta/V$) 
\bea
\Phi_d&=&\frac{2\pi G\Sigma_{da}}{\delta/V}\left(\ln{r}+\ln{(1+\cos{\theta})}\right)\nonumber\\
&+&\Phi^{(r)}_{do}\ln{r}+\Psi_{dr}(\kappa,\theta,r),\label{eq:Phid}
\eea
where $\Sigma_{da}$ is the axi-symmetric surface density so that the first two terms in this potential comprise $\Phi_{da}$. We will use $\Sigma_{dr}$ for the rotating, non-axi-symmetric component. Both $\Sigma_{da}$ and $\Sigma_{dr}$ will include an isothermal gaseous component in general, and so indeed may the volume density $\rho_h$. We adopt the notation $\Phi_{do}\equiv \Phi^{(r)}_{do}+(2\pi G\Sigma_{da}V/\delta)\equiv \Phi^{(r)}_{do}+\Phi^{(a)}_{do} $  subsequently. We note that each of $\Phi_{da}$ and $\Phi_{dr}$ satisfy a Laplace equation above the plane.  

To the extent that the spiral structure is a small component of the disc-halo, we might expect the first approximation for the halo potential to be an axisymmetric function $\Phi_{ha}(\theta)$. This may be written as $\Phi_{ha}=\Phi^{(a)}_{ho}\ln{r}+\Psi_{ha}(\theta)$. The combined solution for $\Psi_a\equiv\Psi_{ha}(\theta)+\Psi_{da}(\theta)$ is given by equation (\ref{eq:solisoDH}). Hence $\Psi_{ha}(\theta)$ may be found by subtraction as 
\bea
&~&\Psi_{ha}=\Psi_a-\Psi_{da}\label{eq:Psiha}\\
&=&\Phi_{oa}\ln{(1+\sqrt{S}\cos{\theta})}-\Phi^{(a)}_{do}\ln{(1+\cos{\theta})}\nonumber
\eea
We recall that $S\equiv (\Phi^{(a)}_{do}/\Phi_{oa})^2$ in current notation, so that $\partial_\theta\Psi_{ha}|_{\pi/2}=0$ as it should. The factor $\Phi_{oa}\equiv \Phi^{(a)}_{do}+\Phi^{(a)}_{ho}$ in current notation.

However the rotating spiral disc structure will impose a rotating perturbation on the halo potential/density $\Phi_{hr}(\kappa,\theta,r)=\Psi_{hr}(\kappa,\theta,r)+\Phi^{(r)}_{ho}\ln{r}$, which form is derived below. It may be  that the  the causal order is inverted. This would mean that the `spiral' halo structure is actually the origin of the disc spiral structure, being itself due for example to the decaying orbit of a merging object. We do not have to decide this point here, as the formalism is the same in either event.  

Consequently we write for the halo potential $\Phi_h\equiv \Phi_{ha}+\Phi_{hr}$
\be
\Phi_{h}=\Phi_{ho}\ln{r}+\Psi_{hr}(\kappa,\theta,r)+\Psi_{ha}(\theta),
\ee
where $\Phi_{ho}\equiv \Phi^{(a)}_{ho}+\Phi^{(r)}_{ho}$. The Poisson equation for the total potential becomes (after adding the Laplace equations for $\Phi_{da}$ and $\Phi_{dr}$ and considering $\theta<\pi/2$)
\bea
&~&\!\!\!\!\!\!\!\!\!\!\!\!\!\!\nabla^2(\Phi_{ha}(\theta)+\Phi_{da}(\theta))+\nabla^2(\Phi_{dr}(\kappa,\theta,r)+\Phi_{hr}(\kappa,\theta,r))\nonumber\\
&=&\frac{Q\Phi_{o}}{r^2}e^{-2(\frac{\Psi_{ha}+\Psi_{da}}{\Phi_o})}~~e^{-2(\frac{\Psi_{dr}+\Psi_{hr}}{\Phi_o})})\label{eq:Pa+r}
\eea
Here $Q$ is as in equation (\ref{eq:Q}) except that $\delta\leftarrow \delta/V$, and $\Phi_o\rightarrow \Phi_{oa}\equiv\Phi_{do}+\Phi_{ho}$. 
 
The axisymmetric part of this last equation (using $\Phi_a\equiv \Phi_{da}+\Phi_{ha}$ and $\Phi_{oa}$ in  $Q$) satisfies 
\be
\nabla^2(\Phi_a)=\frac{Q\Phi_{oa}}{r^2}~~e^{-2\left(\frac{\Psi_a}{\Phi_{oa}}\right)},\label{eq:Pha}
\ee  
and has the solution (\ref{eq:solisoDH}) for $\Psi_a$ when the disc boundary condition is imposed. As remarked above this solution is the simplest approximation to the halo potential, which follows by setting the rotating potential components of the disc and halo equal to zero. The next approximation may be found by expanding the second exponential in equation (\ref{eq:Pa+r}), by neglecting $\Phi_{or}=\Phi^{(r)}_{h0}+\Phi^{(r)}_{do}$ in $\Phi_{oa}$, and by subsequently using equation (\ref{eq:Pha}), to find the inhomogeneous linear equation 
\be
\nabla^2(\Phi_{hr}+\Phi_{dr})=-2\frac{Q}{r^2}e^{-2\left(\frac{\Psi_a}{\Phi_{oa}}\right)}(\Psi_{hr}+\Psi_{dr}).\label{eq:Phr}
\ee 
In this equation $\Psi_{dr}$ is known from equation (\ref{eq:rotPoiss}), together with the disc boundary condition in terms of $\Sigma_{dr}$. The exponential is known from equation (\ref{eq:solisoDH}) with $Q+S=1$.

Our task is now to describe the collisionless material comprising the rotating isothermal halo component. This argument parallels our discussion for the rotating disc component in the previous section, but must be done in spherical geometry with time dependence. Although it is cumbersome, we state here the complete CBE for such a problem. The reduction to the disc CBE is immediate by setting the 3D DF $f=F\delta_D(\theta-\pi/2)\delta_D(v_\theta)$ ($\delta_D$ is the Dirac function) and integating from $\pi/2-\epsilon$ to $\pi/2+\epsilon$ over theta and from $-\epsilon$ to $\epsilon$ over $v_\theta$, and letting $\epsilon\rightarrow 0$. The equation is (recall that $\Omega\equiv V/r$)  in the locally rotating frame 
\bea
&~&\!\!\!\!\!\!\!\!\partial_tf +v_r\partial_rf+\frac{v_\theta}{r}\partial_\theta f+\left(\frac{v_\phi}{r\sin{\theta}}-v_rt\partial_r\Omega\right)\partial_\phi f \nonumber\\
&+&\!\!\!\!\!\!\left(\frac{v^2_\theta+v^2_\phi}{r}+2\Omega\sin{(\theta)}v_\phi+\Omega^2r\sin^2{(\theta)}-\partial_r\Phi_{hr}\right)\partial_{v_r}f\nonumber \\
&+&\!\!\!\!\!\!\!\!\left(\frac{v^2_\phi}{r}\cot{\theta}+2\Omega v_\phi\cos{\theta}+\Omega^2r\sin{(\theta)}\cos{(\theta)}-\frac{v_rv_\theta}{r}-\frac{1}{r}\partial_\theta\Phi_{hr}\right)\nonumber \\
&\times&\partial_{v_\theta}f\nonumber\\
&-&\!\!\!\!\!\!\left(2\Omega v_\theta\cos{\theta}+\frac{v_\phi v_\theta}{r}\cot{\theta}+\frac{v_rv_\phi}{r}+\Omega v_r\sin{\theta}+\frac{1}{r\sin{\theta}}\partial_\phi\Phi_{hr}\right)\nonumber\\
&\times& \partial_{v_\phi}f=0.\label{eq:MsrCBE}
\eea
 
We convert this equation to self-similar variables in the usual way by assigning (the $T$ dependence is because of the gradual destruction of the self-similarity by winding) 
\bea
&~&\!\!\!\!\!\!\!\!\!\!\delta t=e^{\delta T},r=Re^{\delta T},\xi=\phi+\epsilon T,\vec{v}\equiv\vec{Y},\nonumber\\
&~&\!\!\!\!\!\!\!\!\!\!f= P(R,\xi,\theta,\vec{Y};T)e^{-2\delta T},\nonumber\\
&~&\!\!\!\!\!\!\!\!\!\!\Phi_{hr}=\nonumber\\
&~&\!\!\!\!\!\!\!\!\!\!\Phi^{(r)}_{ho}(\ln{R}+\delta T)+\Psi_{hr}(R,\xi,\theta;T),\nonumber\\
&~&\!\!\!\!\!\!\!\!\!\!\rho=\Theta(R,\xi,\theta;T)e^{-2\delta T},\Theta=\int~P~d^3Y.\label{eq:3Dvars}
\eea

The equation that results from inserting these variables into equation (\ref{eq:MsrCBE}) has the following characteristics:
\bea
\frac{dP}{dT}&=&2\delta P,~~\frac{dR}{dT}=Y_R-\delta R,~~\frac{d\theta}{dT}=\frac{Y_\theta}{R},\nonumber\\
&~&\frac{d\xi}{dT}=\epsilon+\frac{Y_\phi}{R\sin{\theta}}+\frac{V}{\delta R}\frac{Y_R}{R},\nonumber\\
\frac{dY_R}{dT}&=& \frac{Y^2_\theta+Y^2_\phi}{2}+\frac{2V}{R}Y_\phi\sin{\theta}+\frac{V^2}{R}\sin^2{\theta}-\partial_R\Phi_{hr},\nonumber\\
\frac{dY_\theta}{dT}&=& \frac{Y^2_\phi}{R}\cot{\theta}+\frac{2V}{R}Y_\phi\cos{\theta}+\frac{V^2}{R}\sin{(\theta)}\cos{(\theta)}\nonumber\\
&-&\frac{Y_RY_\theta}{R}-\frac{1}{R}\partial_\theta\Phi_{hr},\label{eq:halotempchars}\\
\frac{dY_\phi}{dT}&=& -\frac{2VY_\theta}{R}\cos{\theta}-\frac{Y_\phi Y_\theta}{R}\cot{\theta}-\frac{Y_RY_\phi}{R}\nonumber\\
&-&\frac{VY_R}{R}\sin{\theta}-\frac{1}{R\sin{\theta}}\partial_\xi\Phi_{hr}.\nonumber
\eea

We can combine the $R$, $\theta$ and $Y_\phi$ characteristics to obtain 
\bea
&~&\frac{d}{dT}(R\sin{(\theta)}Y_\phi+VR\sin^2(\theta))=\nonumber\\
&-&\delta R(Y_\phi\sin{\theta}+V\sin^2{\theta})-\partial_\xi\Phi_{hr},\label{eq:SSangmom3D}
\eea
which, by returning to physical coordinates becomes the angular momentum equation 
\be
\frac{d}{dT}(r\sin{(\theta)}v_\phi+Vr\sin^2{(\theta)})=-\partial_\xi\Phi_{hr}.\label{eq:3Dangmom}
\ee

Just as is the case for the disc we can combine these characteristics to obtain a relation for the change in the inertial frame  energy along a trajectory  as
\bea
\frac{d}{dT}E_{hr}&=&\partial_T\Phi_{hr}-\delta R\partial_R\Phi_{hr}\nonumber\\
&+&\left(\epsilon+\frac{V}{\delta R}\frac{d\ln{R}}{dT}\right)\partial_\xi\Phi_{hr}.\label{eq:3DIengder}
\eea
Here 
\be
E_{hr}=E'_{hr}+V\sin{(\theta)}(Y_\phi+\frac{V}{2}\sin{\theta}),\label{eq:3Deng}
\ee
where
the energy in the locally co-moving frame is 
\be
E'_{hr}=\frac{\vec{Y}^2}{2}+\Phi_{hr}.\label{eq:3Droteng}
\ee

We substitute the form of $\Phi_{hr}$ from equation (\ref{eq:3Dvars}) into the right-hand side of equation (\ref{eq:3DIengder}), and in order to obtain the conservation of inertial energy we set the resulting expression to zero. This yields the compatible  form of $\Psi_{hr}$, namely $\Psi_{hr}(r,\theta,\xi-\epsilon T+V/(\delta R))$. As for the rotating disc component, we can incorporate the dependence on a transient logarithmic spiral by introducing the variable $\kappa$ to write   
\be
\Psi_{hr}=\Psi_{hr}(r,\theta,\kappa).\label{eq:Psirot3D}
\ee
This justifies the form of the rotating halo potential that we used at the beginning of this section. An explicit dependence on $r=Re^{\delta T}$ destroys the self-similarity.

The DF for this halo component follows from $P=fe^{2\delta T}$ where $f$ can only depend on integral constants. To describe a structure in net rotation we wish it to be a function of $E'_{hr}$. However this energy is only constant according to equation (\ref{eq:3Deng}) for $Y_\phi<V$ and $d\theta/dT$ small. The 'small' must be with respect to $\Omega$ so that using the theta characteristic we require $(d\theta/dT)/\Omega\equiv Y_\theta/V<1$. Thus for consistency the DF must decline rapidly when $Y_\phi$ and $Y_\theta$ exceed $V$. 

The isothermal self-similar form is clearly necessary as 
\be
f=K_{hr}e^{-(\frac{2E'_{hr}}{\Phi^{(r)}_{ho}})},\label{eq:3DDF}
\ee
 since it succeeds in producing the expected $\rho\propto 1/r^2$ with no explicit dependence of $P$ on $T$. This self-similarity is broken through the dependence on $r=Re^{\delta T}$ in $\Psi_{hr}$, as may be required by the Poisson equation.
When calculating the collisionless halo density from the isothermal DF, the total potential must be used and $\Phi^{(r)}_{ho}\leftarrow \Phi_{ho}+\Phi_{do}=\Phi_o$. By our approximations we should have $V^2\ge\Phi_o$.

We turn next to the expressions for the various potential components.

\subsection{Non Axially Symmetric Potential solutions}

The basic potential for the disc-halo system is given by equation (\ref{eq:solisoDH}) for the axi-symmetric component, and the solution of equation (\ref{eq:rotPoiss}) for the spiral disc component. After solving for the disc spiral component, the non-axially-symmetric halo component may be found in principle from equation (\ref{eq:Pa+r}), or approximately from equation (\ref{eq:Phr}). 

However equation (\ref{eq:Phr}) is only readily separable using an ans\"atz of the form  (\ref{eq:Psid}), if one takes $\Phi^{(r)}_o\equiv \Phi^{(r)}_{ho}+\Phi^{(r)}_{do}=0$ and absorbs the log term into the axi-symmetric potential. Equation (\ref{eq:Phr}) then takes the form
\be
{\cal L}\Psi_r=-\frac{2(1-S)}{(1+\sqrt{S}\cos{\theta})^2}\Psi_r,\label{eq:Psihdr} 
\ee
where $\Psi_r\equiv \Psi_{dr}+\Psi_{hr}$. This equation can be solved in separated form, but it is essentially only known as a series. We reserve the complete exploration of such halo spiral structure to another work.

We can only expect to find regular spiral structure in the disc when $Vt/r<\epsilon/\delta$, so that our discussion of regular spiral arms will be restricted to that limit. However to identify a possible description of the evolution of the arms we proceed briefly with the general case.

Equation (\ref{eq:rotPoiss}) with the modal ans\"atz of equation (\ref{eq:Psid}) is resolved into two equations ($\epsilon\leftarrow \epsilon/\delta$)
\bea
&~&\!\!\!\!\!\!\!\!\!\!\!\!\!\frac{1}{\sin{\theta}}\frac{d}{d\theta}(\sin{\theta}\frac{dT}{d\theta})+T(\theta)(k_m^2-m^2(\epsilon^2+\frac{1}{\sin^2{\theta}})+im\epsilon)\nonumber\\
&=&0,\label{eq:sep1}\\
&~&\!\!\!\!\!\!\!\!\!\!\!\!\!\zeta^2\frac{d^2{\cal R}}{d\zeta^2}-im\zeta(\epsilon-\zeta)\frac{d{\cal R}}{d\zeta}-{\cal R}(m^2(\epsilon-\zeta)^2+k_m^2-m^2\epsilon^2)\nonumber\\
&=&0,\label{eq:sep2}
\eea  
where $k_m^2$ is the separation constant (positive or negative in general) and $\zeta\equiv Vt/r$. Near $\zeta=0$ the appropriate solution of the second equation is ${\cal R}=1$ and $k_m^2=0$. The solution to the first equation is then 
simply found in terms of associated legendre functions and so by (\ref{eq:Psid})
\bea
&~&\Phi^m_{dr}(\kappa,\theta)-\Phi^{(r)}_{do}\ln{\sin{\theta}}\equiv \Psi^m_{dr}\nonumber\\
&=&\!\!\!\!\!e^{(im\kappa)}(C_{1m}~P^m_{im\epsilon}(x)+C_{2m}~Q^m_{im\epsilon}(x)),\label{eq:initspiral}
\eea
where $P^\mu_\nu$ and $Q^\mu_\nu$ denote the associated Legendre functions and $C_{1m}$, $C_{2m}$ are  complex modal constants. This potential, together with the axially symmetric disc-halo potential, will be our principal concern below, but it is of some interest to examine the evolving radial dependence.

The modal solution for the radial dependence takes the form  
\bea
{\cal R}_m(\zeta)&=&\exp{i(-\frac{m\zeta}{2}+\frac{m\epsilon}{2}\ln{\zeta})}(A_{1m}M_{\lambda,\mu}(\sqrt{3}m\zeta)\nonumber\\
&+&A_{2m}W_{\lambda,\mu}(\sqrt{3}m\zeta)),\label{eq:asymrad1}
\eea
where $M,W$ are Whittaker functions, $\lambda\equiv \sqrt{3}m\epsilon/2$, and $\mu\equiv \sqrt{(1+im\epsilon)^2+4k_m^2}/2$. 

We can simplify this expression somewhat by considering the radial dependence near $\zeta=\epsilon$, where it should represent a rapidly winding spiral. In this limit the radial equation becomes approximately $d^2{\cal R}/d\zeta^2=(k_m^2/\epsilon^2-m^2){\cal R}$. Hence $k_m^2\le m^2\epsilon^2$ implies an oscillation in $\zeta$ (i.e. $1/r$ at fixed time) corresponding to a winding of the arm. The value $k_m^2=\epsilon^2m^2$ corresponds to a marginally stable case, where the deviation  from an arm intially constant on $\kappa=constant$ is linear in $1/r$. In this limit $\mu=\sqrt{(3m\epsilon-i)(m\epsilon+i)/2}$ in the Whittaker functions. 

The asymptotic behaviour of the Whittaker functions at large argument are 
\bea
M_{\lambda,\mu}&\asymp& \frac{\Gamma(1+2\mu)}{\Gamma(1/2+\mu-\lambda)}\frac{e^{\sqrt{3}m\zeta/2}}{(\sqrt{3}m\zeta)^\lambda},\nonumber\\
W_{\lambda,\mu}&\asymp& \frac{e^{-\sqrt{3}m\zeta/2}}{(\sqrt{3}m\zeta)^\lambda}.
\eea
 Consequently, it is the Whittaker M function that describes the destruction of the spiral arm with increasing $\zeta$. It is of interest that this destructive evolution produces an oscillating exponential decreasing with increasing radius along the arm. The amplitude is proportional to $(r/mVt)^\lambda e^{(mVt/r)\sqrt{3}/2}$

The modal analysis for the function $\Psi_r$ based on equation (\ref{eq:Psihdr}) yields the two equations (\ref{eq:sep1}, \ref{eq:sep2}); but with the additional term $2(1-S)/(1+\sqrt{S}\cos{\theta})^2$ in the bracket multiplying $T(\theta)$, in the first of these equations. The resulting equation is solvable formally in terms of a Heun series, but it is best studied numerically. A non-trivial exception is when the disc dominates the halo so that $S\approx 1$. Then the halo spiral structure  satisfies the homogeneous equation (\ref{eq:rotPoiss}), but with different boundary conditions. A sum over modes might be required to describe the orbit of an infalling object, but the separated modal form of equations (\ref{eq:sep1}, \ref{eq:sep2}) is relevant if the disc spiral is the origin of the halo disturbance. 

We are now equipped to consider in the next section the properties of transient spiral arms embedded in an axi-symmetric, isothermal, disc-halo.

\section{Transient  Spiral Arms in an Isothermal Disc-Halo System}

We study in this section examples of `initial' (the creation of the disturbance does not concern us here) spiral arms in a disc-halo system. Various components are considered. These are comprised  of collisionless particles with the corresponding isothermal DF, and/or isothermal gas. There is the axi-symmetric disc-halo background, the spiral disturbance in the disc associated with the spiral arms themselves and the consequent spiral disturbance in the halo. The latter component does not appear in the following disc boundary condition, but it could be observable in edge-on galaxies. Its simplest form would be a series of logarithmic spirals on cones with amplitude decreasing with decreasing $\theta$. 

Collisional material must be present, in order to satisfy everywhere the boundary condition 
\be
\sigma_a+\sigma_{dr}+\sigma_g=-\frac{1}{2\pi Gr}(\partial_\theta\Psi_a|_{\pi/2}+\partial_\theta\Psi_{dr}|_{\pi/2}).\label{eq:bctot}
\ee
Here $\sigma_a$ is the axi-symmetric background density, $\sigma_{dr}$ is the rotating spiral density, and $\sigma_g$ is an isothermal gas density.
We can calculate $\sigma_a$ and $\sigma_{dr}$ from their corresponding distribution functions in terms of the potential components that were presented in the last section. The gas density is taken normally to be determined by this boundary condition. 

In fact the gas distribution is subject to the same potentials as are the other disc  
components and in principle its velocity and density are determined by the hydrodynamic equations. However the gas behaviour is subject to the magnetic field, particularly in the inter-arm regions. Thus, in the absence of major streaming, the gas is likely to be in (isothermal) magneto-hydrodynamic (MHD) quasi-equilbrium. By fixing the gas density from the boundary condition (\ref{eq:bctot}), we are effectively determining the (quasi, because of the winding spiral arms) quasi- equilibrium magnetic field. Such a field in the disc would have the equilibrium form $\vec{B}=\vec{b}e^{-\delta R}$, where $\vec{b}=\vec{b}(\kappa,r)$. This raises the possibility of comparing the consequent magnetic field structure with observations. But we  leave this aspect to a future work as the required MHD equations are formidable. An eventual complete disc solution will require this gap to be closed.

The boundary condition (\ref{eq:bctot}) reduces to 
\be
\sigma_{dr}+\sigma_{gr}=-\frac{1}{2\pi Gr}(\partial_\theta\Psi_{dr}|_{\pi/2}),\label{eq:bcrotgas}
\ee
 when we recall that $\Psi_a$ is defined so that 
\be
\sigma_a\equiv -\frac{1}{2\pi G r}(\partial_\theta\Psi_a|_{\pi/2}).\label{eq:bcaxi}
\ee
Because we take $\Psi_a(\pi/2)=0$, and because we may choose $\Psi_{hr}(\pi/2)=0$ as well as absorbing $\Phi^{(r)}_{do}$ and $\Phi^{(r)}_{ho}$ into $\Phi_{oa}$, we can write
\be
\sigma_{dr}=2\pi \Phi_{oa}\frac{K_{dr}}{r}e^{-\frac{\Psi_{dr}}{\Phi_{oa}}}.\label{eq:rotdens2}
\ee
For the gas density we take $\sigma_{gr}=\Sigma_{gr}(\kappa,r)/(\delta r)$ where for the initial spiral structure the dependence on $r$ is ignorable, just as for the potential.
  
We begin our investigation of the boundary condition (\ref{eq:bcrotgas}) by considering the solution (\ref{eq:initspiral}) for the initial spiral disc potential in more detail.

 We  only consider one mode at a time in this treatment, normally $m=2$ since this correponds to many observed spirals. One might retain a full fourier analysis of $\Phi_{dr}$ in order to satisfy the boundary condition (\ref{eq:bcrotgas}) over a range of $\kappa$, but one mode would still have to be dominant in order to match the observations.

The constants in the solution (\ref{eq:initspiral}) may  be chosen freely, but it is useful to check that our symmetry requirement $\Psi^m_{dr}(\theta)=\Psi^m_{dr}(\pi-\theta)$ is satisfied . If we consider $\Psi^m_{dr}$ to increase away from the disc towards the axis, then this symmetry enforces our assumed asymmetric boundary condition $(\partial_\theta\Psi^m_{dr})_{\pi/2-}=-(\partial_\theta\Psi^m_{dr})_{\pi/2+})$. Moreover the gravitational acceleration of the disc is then towards the disc.

 The Legendre functions at positive $x$, using the `cut' employed for example in \cite{GR94}, have a non trivial relation to those at negative $x$. To enforce the symmetry, and to retain two free constants, the constants $C_{1m}$ and $C_{2m}$ must be related to the constants at $x=0-$, namely $C_{1m}^-,~C_{2m}^-$, by the relations
\bea
C_{1m}^-&=&C_{1m}\cos{\phi_c}-\frac{\pi}{2}~\sin{\phi_c}~C_{2m},\nonumber\\
C_{2m}^-&=&-\frac{2}{\pi}~\sin{\phi_c}~C_{1m}-\cos{\phi_c}C_{2m}.\label{eq:constcon1}
\eea 
 We have defined the complex angle $\phi_c\equiv (1+i\epsilon)m\pi$. 

It is possible to insist that the constants retain their values across the disc because equations (\ref{eq:constcon1}) then become homogeneous with a zero determinant. However in such a case the ratio of the constants is defined in the form 
\be\frac{C_{2m}}{C_{1m}}=-\frac{2}{\pi}~\frac{\sin{\phi_c}}{1+\cos{\phi_c}}=-\frac{2i}{\pi}~\frac{\sinh{(m\epsilon\pi)}}{(-1)^m+\cosh{(m\epsilon\pi)}}.\label{eq:constcon2}
\ee
 Once the winding angle of the spiral disturbance is fixed, this condition reduces the free constants to one. Moreover we can suppose that $C_{1m}$ is real since any phase constant will simply add an arbitrary phase to $e^{im\kappa}$. However this reduced case may not allow us to have spiral arms of arbitrary amplitude.

We have taken the $\kappa$ dependence of our mode to be periodic, as is customary in linear wave descriptions. However in a non-linear treatment aperiodic solutions may also be possible. These would require $m=-ip$ where $p$ is a real number, and so the potential would be aperiodic. This implies discontinuities in the spiral disc potential. In a non-linear disturbance after coarse graining such discontinuities may be realized as collisionless `shocks'. In resolved detail they would be regions of rapidly changing potential and surface density, probably involving normal gas shocks. They might be expected along the  edges of the spiral arms, but we shall not consider this possibility further in this paper.  

The boundary condition (\ref{eq:bcrotgas})becomes explicitly ($x=\cos{\theta}$)
\bea
&~&\Sigma^m_{gr}(\kappa)+2\pi \Phi_{oa}K^m_{dr}\exp{(-(\frac{\Psi^m_{dr}(0)}{\Phi_{oa}}))}\nonumber\\
&=&\frac{\delta}{2\pi G V}\partial_x\Psi^m_{dr}|_{0},\label{eq:bcexplicit}
\eea

where $\Psi^m_{dr}(0)$ is given by the real part of equation (\ref{eq:initspiral}) and the real part of the derivative is also known from this expression. We have reintroduced units of length here so that $G$ may have its normal dimensions and value.

This last equation can be satisfied everywhere by taking it to be an equation for $\Sigma^m_{gr}(\kappa)$,provided that the constants may be chosen so that the net gas density $\Sigma_{ga}+\Sigma^m_{gr}$ ($\Sigma_a$ is re-labelled $\Sigma_{ga}$ for consistency) is always positive.   

We take as an example the asymmetric mode $m=2$ with an initial winding angle $\epsilon/\delta=2$, which corresponds to the complementary pitch angle of $26^\circ.56$. Then the boundary condition becomes (dropping the $m=2$ superscript)
\bea
\Sigma_{gr}(\kappa)&=&\frac{\delta}{2\pi G V}[\Phi_{oa}y_1DP\cos{(\phi_{DP}+\phi_{12}+2\kappa)}\nonumber\\
&+&\Phi_{oa}y_2DQ\cos{(\phi_{DQ}+\phi_{22}+2\kappa)}]\nonumber\\
&=&-2\pi \Phi_{oa}K_{dr}\exp{[-y_1P\cos{(\phi_P+\phi_{12}+2\kappa)}]}\nonumber\\
&\times&exp{[-y_2Q\cos{(\phi_Q+\phi_{22}+2\kappa)}]},\label{eq:mastercon}
\eea
where we have introduced the `ad hoc' definitions
\bea
P^2_{4i}(0)&=& \!\!\!\!\!\!\!\!\!\!\frac{4}{\sqrt{\pi}}\cos{(\pi(1+2i))}\frac{\Gamma(3/2+2i)}{\Gamma(2i)}\equiv Pe^{i\phi_P},\nonumber\\
Q^2_{4i}(0)&=& \!\!\!\!\!\!\!\!\!\!-2\sqrt{\pi}\sin{(\pi(1+2i))}\frac{\Gamma(3/2+2i)}{\Gamma(2i)}\equiv Qe^{i\phi_Q}, \nonumber\\
\partial_x(P^2_{4i})|_0&=& \!\!\!\!\!\!\!\!\!\!\frac{8}{\sqrt{\pi}}\sin{(\pi(1+2i))}\frac{\Gamma(2(1+i))}{\Gamma(2i-1/2)}\equiv DP~e^{i\phi_{DP}},\nonumber\\
\partial_x(Q^2_{4i}|_0&=&\!\!\!\!\!\!\!\!\!\!4\sqrt{\pi}\cos{(\pi(1+2i))}\frac{\Gamma(2(1+i))}{\Gamma(2i-1/2)}\equiv DQ~e^{i\phi_{DQ}},\nonumber
\eea 
in terms of the phases and absolute values of the associated Legendre functions. In addition $y_1\equiv |C_{12}|/\Phi_{oa}$ and $y_2\equiv |C_{22}|/\Phi_{oa}$. With the moduli and phases of $C_{12}$, $C_{22}$ given, plus the real values for $\Phi_{oa}$ and $K_{dr}$, equation (\ref{eq:mastercon}) determines the required non-axially symmetric gas density. A simpler example is given by equation (\ref{eq:constcon2}), which for the present example gives $C_{22}/C_{11}=-2i/\pi$ very nearly, so that with $\phi_{12}=0$, $\phi_{22}=-\pi/2$ and $y_2/y_1=2/\pi$. This leaves only $\Phi_{oa}$, $K_{dr}$ and $y_1$ to be assigned.

This procedure does have the merit of indicating that, without gas, spiral structure is not possible in this theory. However there are hidden consequences to be explored, since in the axi-symmetric disc-halo the values of $\Sigma_{ga}$, $\Phi_{oa}$, and corresponding halo quantities are all constrained by $S+Q=1$.
Moreover, this approach makes the dynamics of the gas subject to this boundary condition, which dynamics we do not explore further in this work. 

It is however instructive to consider the other extreme, wherein the non-axially symmetric gas behaviour is unimportant. This can be done by taking its value in equation (\ref{eq:mastercon}) to be constant. We  find then, as was indicated earlier, that the one mode boundary condition is only readily satisfied at discrete values of $\kappa$. These discrete values then delineate the `arms'.  Equation (\ref{eq:mastercon}) must then be satisfied with constant $\Sigma_{gr}$ in the arms, while between the arms we consider only  the $\psi_a$,$\Sigma_{ga}$ pair.

 To be consistent we must neglect the effect of $\partial_x\Psi_{dr}|_o$ on the surface density in the inter-arm region. For example let us suppose that the `gas' is in pure rotation with a constant linear speed $V_\phi$, and that it obeys the isothermal self-similarity generally. This requires $\sigma_g=\Sigma_g(\kappa)e^{-\delta R}$, and the `pressure' in the surface due to this component to have the form $p_s=\tau(\kappa)e^{-\delta R}$. Then by applying the radial and azimuthal equations of equilibrium to the gas in the inertial frame, we obtain respectively (prime denotes $d/d\kappa$ and we include neither viscosity nor magnetic field)
\be
\tau'-\frac{\delta}{\epsilon}\tau=\frac{\delta}{\epsilon}\Sigma_{g}(V_\phi^2-\Phi_{oa}),
\ee
and
\be
 \frac{d}{d\kappa}\left(\frac{V_\phi^2}{2}+\Psi_{dr}+\frac{\tau}{\Sigma_{g}}\right)=\frac{\tau}{\Sigma_{g}}\frac{d\ln{\Sigma_{g}}}{d\kappa}.
\ee
By neglecting $\Psi_{dr}$ and in addition by requiring $\Sigma_g=\Sigma_{ga}$ and hence $\tau$ to be constant, we see that a simple solution gives 

\be
\frac{\tau}{\Sigma_{ga}}+V_\phi^2-\Phi_{oa}=0
\ee
 
In general the gas dynamics is  complex, and if we accept the value of the surface density from the boundary condition (\ref{eq:mastercon}), we would  have to solve the gas equations for the magnetic field if the gas is in equilibrium. There may well be a flow field  in addition to the magnetic field and indeed time dependence, all to be rendered consistent with the boundary  induced value. But this problem is decoupled from the problem we solve here.  

For definiteness we study the numerical solution of the boundary condition under the simplifying assumptions of equation (\ref{eq:constcon2}). It happens that, for a two-armed mode with a large winding angle that we study here, the condition of equation (\ref{eq:constcon2}) is very nearly the same as requiring $\Psi_{dr}(\kappa,0)/\Phi_{oa}=0$. There is a difference in the two conditions at the level of a part in $10^{-5}$, which allows $\Psi_{dr}(\kappa,0)$ to be small but not strictly zero. We use this  example here for illustration, but in general $\Psi_{dr}$ may not be as small, since the constants are arbitrary provided equations (\ref{eq:constcon1}) are satisfied.

An inspection of the boundary condition under these conditions confirms that the constants may be chosen so as to satisfy it only at discrete values of $\kappa$. We choose these to be at $\kappa=0$ and at $\kappa=\pi$ for the two-armed case. The boundary condition (\ref{eq:mastercon}) becomes thus at $\kappa=0$ 
\bea
\frac{2\pi GV}{\delta \Phi_{oa}}\Sigma_{gr}&+&\frac{4\pi^2GV}{\delta}K_{dr}\exp{\left[-P\cos{\phi_P}-\frac{2}{\pi}Q\sin{\phi_Q}\right]y_1}\nonumber\\
&=&y_1(DP\cos{\phi_{DP}}+\frac{2}{\pi}DQ\sin{\phi_{DQ}}),
\eea
whichis an equation for $y_1$. This latter measures the strength of the spiral potential relative to the axi-symmetric potential. The parameter $E\equiv (2\pi GV/\delta)(\Sigma_{gr}/\Phi_{oa})$ measures the surface density of gas in the spiral arm while the parameter $A\equiv (4\pi^2GV/\delta)K_{dr}$ is a measure of the collisionless surface density in the arm, both essentially as a fraction of the axi-symmetric surface density. The bracketed expressions that depend on the amplitudes and phases of the associated Legendre functions are pure numbers.

If we write this last expression schematically as 
\be
E+Ae^{-Cy_1}=By_1,
\ee
then $B \equiv DP\cos{\phi_{DP}}+\frac{2}{\pi}DQ\sin{\phi_{DQ}}\approx 2.48\times 10^5$ and $C\equiv -P\cos{\phi_P}-\frac{2}{\pi}Q\sin{\phi_Q} \approx -1.6\times 10^3$. The parts depending on $Q^2_{4i}(0)$ are negligible. 

This schematic relation is easily solved given values for $E$ and $A$. Relatively large choices such as $A=0.5$ and $E=0.05$ {\it or vice versa} give $y_1\approx -2.2\times 10^{-6}$. At $A=0.1$ and $E=0.01$ {\it or vice versa} give $y_1\approx -4.4\times 10^{-7}$. Either choice renders the spiral potential small in terms of the axi-symmetric structure of the disc as is required for inter-arm  consistency.
 
The opposite extreme, where the gas distribution is essential, can be illustrated by solving the boundary condition (\ref{eq:mastercon}) for $\Sigma_{gr}(\kappa)$ under the approximation of equation (\ref{eq:constcon2}). In this procedure we assign $y_1$ as well as the parameters $E$ and $A$.

Thus in terms of a schematic notation we have,
\bea
&~&E(\kappa)=y_1(B\cos{2\kappa}+B1\sin{2\kappa})\nonumber\\
&-&\!\!\!\!\!A\exp{(y_1(-C\cos{2\kappa}+C1\sin{2\kappa}))},\label{eq:schematicgas}
\eea
where $B1\equiv -DP\sin{\phi_{DP}}+\frac{2}{\pi}DQ\cos{\phi_{DQ}}\approx 7.5\times 10^3$ and $C1\equiv P\sin{\phi_P}-\frac{2}{\pi}Q\cos{\phi_Q}\approx 5.3\times 10^4$. This yields the function $E(y_1,A)$ for the scaled spiral gas density that is required as a function of the physical parameters.

\begin{figure}
\begin{tabular}{cc} 
{\rotatebox{0}{\scalebox{.5} 
{\includegraphics{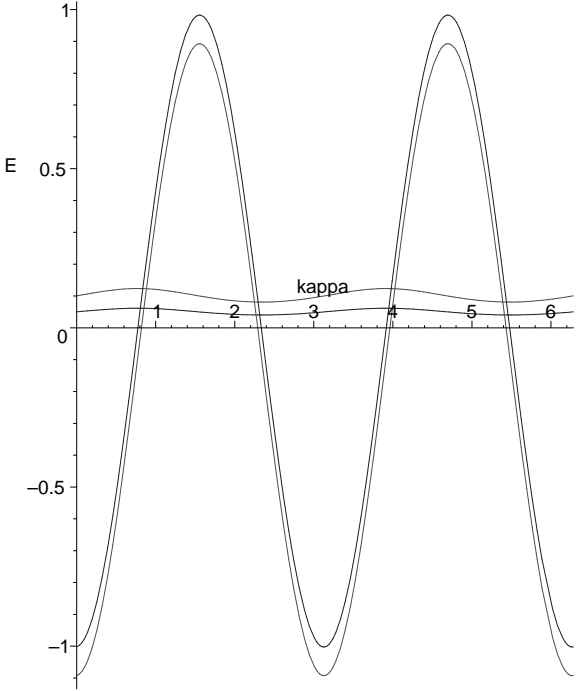}}}}
\end{tabular}
\caption{ The larger pair of curves is the non axially-symmetric gas density as a function of $\kappa$ for $y_1=4\times 10^{-6}$ with $A=0.01$ at top and $A=0.1$ underneath.  The smaller pair of curves give the non axially-symmetric collisionless particle density for the same value of $y_1$ and $A=0.1$ above and five times the curve for $A=0.01$ below. Larger values of $y_1$ will produce negative total gas density. The gas density responds very strongly to the weak spiral potential.  }    
\label{fig:sigmar}
\end{figure}

In figure (\ref{fig:sigmar}) we show two examples of the gas density dependence on $\kappa$ for $A=0.01$ (upper large amplitude curve), $A=0.1$ (lower large amplitude curve) and $y_1=4\times 10^{-6}$. For larger values of $y_1$ the amplitude of the spiral oscillation tends to dominate the axi-symmetric density. Larger values of $A$ render the oscillation more asymmetric about zero, with the minima coming to dominate the axi-symmetric gas density.

The smaller pair of curves on the figure show the collisionless particle density variation for $A=0.1$ (top curve) and $A=0.01$ (bottom curve-that has been multiplied by five for visibility)) for the same value of $y_1$ as for the gas density. We see that  the relatively small variation in the particle density (and potential) leads to a magnified reaction in the gas density variation by more than a factor ten. The gas density peak is slightly leading (larger $\phi$) the particle density peak at a given radius and is slightly outside (larger $\ln{r}$. This reverses as the winding proceeds, as can be seen from figure (\ref{fig:windspiral})  

This concludes our model for transient spiral arms in the thin disc limit. Many variations of the model are possible if $C_2$ is decoupled from $C_1$. The principal characteristic of this model is that the spiral arm is co-moving with the background disc until it is destroyed by winding. It is perhaps worth remarking also that an oscillating exponential decline appears in the spiral potential
as the arm is wound up, which would lead to similar transient behaviour in the surface density of the disc.

\subsection {Summary of the  Distribution Function}
 
The various components of our disc-halo system have been described in terms of their individual distribution functions. The question arises as to whether the sum of these distribution functions is `valid' (i.e. satisfies the CBE) description of the whole system. Fortunately, under the approximation that in the locally rotating frame $v_\phi<V$, the axi-symmetric and non axi-symmetric distribution functions take the same isothermal form. This holds for the disc and for the halo. Hence writing the appropriate isothermal DF with the total potential, remains a solution of the CBE for every component.  

The sum DF is comprised of  two halo components and two disc components. If we use the approximation wherein the comoving $v_\phi<V$ (the exact DF is the disc part of equation (\ref{eq:comboDF})) these take the form 
\bea
f&=&K_{ha}e^{-2\frac{E_{ha}}{\Phi_{oa}}}+K_{hr}e^{-2\frac{E'_{hr}}{\Phi_{oa}}}\nonumber\\
&+&K_{da}e^{-\frac{E'_{da}}{\Phi_{oa}}}+K_{dr}e^{-\frac{E'_{dr}}{\Phi_{oa}}},\label{eq:DFtot}
\eea
where (all velocities are in the comoving frame)
\bea
E_{ha}&=&\frac{v_r^2+(v_\phi+V)^2+v_\theta^2}{2}+\Phi,\nonumber\\
E'_{hr}&=& \frac{v_r^2+v_\phi^2+v_\theta^2}{2}+\Phi,\\
E'_{da}&=& \frac{v_r^2+v_\phi^2}{2}+\Phi(\pi/2),\nonumber\\
E'_{dr}&=& \frac{v_r^2+v_\phi^2}{2}+\Phi(\pi/2).\nonumber
\eea
In these expressions the total potential is 
\be
\Phi=\Phi_{oa}\ln{r}+\Psi_a(\theta)+\Psi_{dr}(\kappa,r)+\Psi_{hr}(\kappa,r),\label{eq:totpot}
\ee
with the $r$ dependence only developing as the winding continues and $\Psi_a(\pi/2)=0=\Psi_{hr}(\pi/2)$. To the extent that $\Psi_{dr}$ and $\Psi_{hr}$ are small compared to $\Phi_{oa}$ they might be neglected for the collisionless particles. However we have seen that these small potentials can have a major influence on the gas distribution. Moreover they are essential to the spiral nature of the disc and halo.
  
The distribution function approach that we have used for each component avoids the question of the actual particle orbits. These may be found in principle from the characteristic equations of the non axi-symmetric Boltzmann equations for the disc (\ref{eq:timedepchars}) and for the halo (\ref{eq:halotempchars}). The initial state orbits, before significant winding, can be studied by neglecting the terms in $Vt/r$. The halo spiral density disturbance can be found in principle by using $\Phi_{hr}$ in the Poisson equation 
\be
\nabla^2\Phi_{hr}=4\pi G\rho_{hr}.
\ee  
The detailed study of the corresponding orbits must await another work, but it is clear that the resulting spiral distortions  are of interest as possible infalling orbits.

\section{Discussion and Conclusions}

We have studied the construction of spiral arms and discs based on the distribution functions that are dictated largely by isothermal self-similarity.  Section (2.1) incorporates the axially symmetric Mestel disc into this scaling class, by using the frame with constant rotational velocity . This discussion leads to section (2.2) where the compatible (collisionless) halo is studied in some detail in the inertial frame. The solution for the isothermal disc-halo potential is given in equation (\ref{eq:solisoDH}) and the combined disc-halo distribution function is given in equation (\ref{eq:comboDF}). It implies an approximate integral in terms of the energy normal to the disc. this solution forms the background for the spiral structure.

 Most of our new results are to be found in section (3). Here we treat a  spiral arm that is comoving with the isothermal background disc. Consequently it is subject to secular winding  and hence is transient. By treating isothermal self-similarity in a time dependent fashion, we were able to show the effect of the winding on the distribution function and on its potential. This winding may be neglected up to a certain time at a certain radius, which time increases directly with radius. Thus the transient arm is perturbed from the inside out. 

Beyond a critical radius at a given time, the distribution function remains isothermal. The potential is required to be a function only of the spiral coordinate $\kappa$ as an initial condition, but it becomes progressively dependent  on radius as the winding destroys the self-similarity of the arm. As the destruction proceeds the non-axially symmetric potential adopts an oscillating exponential behaviour. This can be significant in the gas distribution by non-linear amplification. If so then after many episodes of transient spirals the isothermal disc will become exponential, although not in the Sersic form.

Another unusual element of the model is the necessary, isothermal  non axially symmetric structure in the halo. This might be observable in edge-on spiral galaxies as a faint symmetric thickening of the disc due entirely to the disc arms.  
 
In section (4) we construct an example of the scale-free, isothermal, disc-halo
 spiral system. The initial arms are rigorously discrete in the case of constant gas density, but the more likely case has the gas surface density reacting strongly to the spiral potential. Examples of this latter variation have been given in figure (\ref{fig:sigmar}). To establish full consistency of the model, the magnetohydrodynamics of the  gas should be studied in the presence of the system potential. This is rather a complicated proposition as it is likely that the isothermality is broken in reality by the physics of heating and cooling. Moreover the sources of the magnetic field are uncertain.
 
 There is much left undone at this stage even for the isothermal class. This includes discussing the possible origin of isothermality during galactic formation and evolution. This is equivalent to asking for the origin of the $a=1$ self-similarity, which does seem to arise naturally in certain regions of simulated dark matter halos. However this theory and the recent simulations do seem to agree on a new picture for at least repeatedly excited spiral arms.

\section{Acknowledgements}

Queen's University at Kingston is to be thanked for their partial support of this research.

\label{lastpage}


\begin{thebibliography}{Subramanian {\em et al.} 99a}
\bibitem[Binney \&Tremaine (2008)]{BT08} Binney, J. \& Tremaine, S., 2008, \textbf{Galactic Dynamics, 2nd Ed.}, \emph{Princeton University Press}
\bibitem[Broeils\&Rhee (1997)]{BR97} Broeils, A.H. \& Rhee, M-J., 1997, Ap.J. \textbf{324},\emph{877}
\bibitem[Carter \& Henriksen (1991)]{CH91} Carter,B.\& Henriksen, R.N., 1991, J. Math. Phys. \textbf{32},\emph{2580}
\bibitem[Einasto (2011)]{E11} Einasto, J., 2011, arxiv:1109.5580
\bibitem[Evans (1993)]{E93} Evans, N. W., 1993, MNRAS \textbf{260}, \emph{191}
\bibitem[Evans (1994)]{E94} Evans, N.W.,1994, MNRAS \textbf{267},\emph{333}
\bibitem[Evans\&Read (1998a)]{ER98a} Evans,N.W.\& Read, J.C.A.,1998, MNRAS \textbf{300},\emph{83}
\bibitem[Evans\&Read (1998b)]{ER98b} Evans,N.W.\& Read, J.C.A.,1998, MNRAS \textbf{300},\emph{106}
\bibitem[Freeman (1970)]{F70} Freeman, K.C., 1970, ApJ \textbf{160},\emph{811}
\bibitem[Foyle et al. (2011)]{FRDLW11} Foyle, K., Rix, H-W, Dobbs, A.K., Leroy, A.K. \& Walter, F., Ap.J. \textbf{735}, \emph{101}
\bibitem[Goodman\&Evans (1999)]{GE99} Goodman, J.\& Evans, N.W.,1999, MNRAS \textbf{309},\emph{599}
\bibitem[Gradshteyn \& Ryzhik (1994)]{GR94} Gradshteyn, I.S. \& Ryzhik, I.M., 1994, Tables of Integrals, Series, and Products; 5th Edition, Alan Jeffrey, Ed., Academic Press, London
\bibitem[Henriksen \& Widrow (1995)]{HW95} Henriksen, R.N. \& Widrow, L.M., 1995,
MNRAS \textbf{276},\emph{679}
\bibitem[Kawata,Grand \& Cropper (2011)]{KGC11} Kawata,D., Grand,R.J.J., \&Cropper, M.,
2011, arxiv:1110.3824v1
\bibitem[Henriksen (2011)]{Harxiv11} Henriksen, R.N., 2011, arxiv:1110.5670
\bibitem[Kalnajs (1976)]{K76} Kalnajs, A.J.,1976, Ap.J. \textbf{205},\emph{751}
\bibitem[Le Delliou, Henriksen \& MacMillan (2011)]{LeDHM11b}Le Delliou, M., Henriksen, R.N. \& MacMillan, J.D., 2011, A\&A \textbf{526}, \emph{A13} 
\bibitem[Lin, C.C. \& Shu, F.H. (1966)]{LS66} Lin,C.C.\& Shu, F.H., Proc. Nat. Acad> Sci.,\textbf{55},\emph{229}
\bibitem[MacMillan,Widrow\& Henriksen (2006)]{MWH06} MacMillan,J. D., Widrow, L.M. \& Henriksen, R.N., 2006, ApJ \textbf{653},\emph{43}
\bibitem[Martos\&Cox (1998)]{MC98} Martos, M.A. \& Cox,D.P., 1998, Ap.J. \textbf{509},\emph{703}
\bibitem[Monet, Richstone \& Schechter (1981)]{MRS81} Monet, D.G., Richstone, D.O. \& Schechter, P.L., 1981, Ap.J. \textbf{245}, \emph{454}
\bibitem[Sellwood (2011)]{JS2011} Sellwood, J., 2011, MNRAS, \textbf{410}, \emph{1637}
\bibitem[Sellwood (2012)]{JS2012} Sellwood, J., 2012, Ap.J., \textbf{751},\emph{44} 
\bibitem[Toomre (1982)]{T82} Toomre, A., 1982, Ap.J. \textbf{259}, \emph{535}
\bibitem[Wada,Baba \& Saitoh (2011)]{WBS11} Wada, K., Baba, J. \& Saitoh, T.R., Ap.J. \textbf{735}, \emph{1}
\end{thebibliography}
 \end{document}